\documentclass{aa}  
\usepackage{xcolor}
\usepackage{graphicx}
\usepackage{rotating}
\usepackage{siunitx}
\usepackage{txfonts}
\usepackage{subfig}
\usepackage{hyperref}
\usepackage{tablefootnote}
\usepackage{threeparttable}
\usepackage[compatibility=false]{caption}
\usepackage{caption}
\hypersetup{
    colorlinks=true,
    citecolor=blue,
    }
\usepackage[normalem]{ulem}    
\usepackage{xcolor}
\usepackage{placeins}    
\usepackage[capitalise]{cleveref}
\begin{document} 
   \title{X-Shooting ULLYSES: Massive stars at low metallicity}
   \subtitle{XIV. Properties of SMC late-O and B supergiants reveal the metallicity dependence of winds in the Magellanic Clouds.}
   
   \author{
          T.\ Alkousa\inst{\ref{inst:sheff}}
          \and
          P.A.\ Crowther\inst{\ref{inst:sheff}}
          \and
          J.M.\ Bestenlehner\inst{\ref{inst:sheff}, \ref{inst:sheff2}}
          \and
          H.\ Sana\inst{\ref{inst:KUL}}
          \and
          F.\ Tramper\inst{\ref{inst:madrid}}
          \and
          J.S.\ Vink\inst{\ref{inst:armagh}}
          \and
          F.\ Najarro\inst{\ref{inst:madrid}}          
          \and
          A.A.C.\ Sander\inst{\ref{inst:ari}, \ref{inst:iwr}}
          \and
          M.\ Bernini-Peron\inst{\ref{inst:ari}}
          \and
          L. Oskinova\inst{\ref{inst:potsdam}}
          \and
          J.Th.\ van Loon\inst{\ref{inst:keele}}
          \and
          R.\ Kuiper\inst{\ref{inst:essen}}
          \and
           The XShootU collaboration
           }
    \institute{
            {Astrophysics Research Cluster, School of Mathematical and Physical Sciences, University of Sheffield, Hicks Building, Hounsfield Road, Sheffield S3 7RH, United Kingdom \label{inst:sheff}}
              \email{Talkousa1@sheffield.ac.uk}
            \and
            {School of Chemical, Materials and Biological Engineering, University of Sheffield, Sir Robert Hadfield Building, Mappin Street, Sheffield S1 3JD, United Kingdom \label{inst:sheff2}}
            \and
            {Institute of Astronomy, KU Leuven, Celestijnenlaan 200D, B-3001, Leuven, Belgium \label{inst:KUL}}
            \and
            {Departamento de Astrof{\'i}sica, Centro de Astrobiolog{\'i}a, (CSIC-INTA), Ctra. Torrej{\'o}n a Ajalvir, km 4, 28850 Torrej{\'o}n de Ardoz, Madrid, Spain \label{inst:madrid}}
            \and   
            {Armagh Observatory and Planetarium, College Hill, Armagh BT61 9DG, United Kingdom \label{inst:armagh}}
            \and
            {Zentrum f{\"u}r Astronomie der Universit{\"a}t Heidelberg, Astronomisches Rechen-Institut, M{\"o}nchhofstr. 12-14, 69120 Heidelberg \label{inst:ari}}
            \and
            {Interdisziplin{\"a}res Zentrum f{\"u}r Wissenschaftliches Rechnen,
            Universit{\"a}t Heidelberg, Im Neuenheimer Feld 225, 69120 Heidelberg,
            Germany\label{inst:iwr}}
            \and
            {Institut f{\"u}r Physik und Astronomie, Universit{\"a}t Potsdam,
            Karl-Liebknecht-Str 24/25, D-14476 Potsdam, Germany \label{inst:potsdam}}
            \and
            {Lennard-Jones Laboratories, Keele University, ST5 5BG, United Kingdom \label{inst:keele}}  
            \and
            {Faculty of Physics, University of Duisburg-Essen, Lotharstra{\ss}e 1, D-47057 Duisburg, Germany \label{inst:essen}} 
            }

   \date{27 October 2025; 28 January 2026}

  \abstract 
   {Hot massive stars lose mass through radiation-driven winds, producing significant chemical, radiative, and mechanical feedback in the surrounding environment. The properties of these winds play a crucial role in determining the star's evolutionary path. Considering the physics of radiation-driven winds, the wind properties should depend on the metal content of the stellar atmosphere. Therefore, studying the wind properties of massive stars in different metallicities (Z) provides a sanity check on prescriptions that are widely used in evolutionary calculations.}
   {We first aim to obtain the stellar and wind properties of a sample of late-O and B supergiants in the Small Magellanic Cloud (SMC). Using these properties, we aim to quantify the dependence of wind properties on metallicity by comparing them with those of a Large Magellanic Cloud (LMC) counterpart study, which has a similar sample and data, and employed the same modelling techniques used in this study.}
   {Spectroscopic modelling of UV and optical data from ULLYSES and XShootU was performed using the radiative transfer code \textsc{CMFGEN}. We also employed an updated Bayesian inference method similar to \textsc{BONNSAI} to explore the evolutionary history of our sample.}
   {We derived the stellar and wind properties of 20 late-O and B supergiants. We derived the following metallicity-dependent recipe for wind momentum: $\log{D_{\rm mom}} = (1.64-0.75\log{Z/Z_{\odot}}) \log{(L_{\rm bol}/10^{6}L_{\odot})} + 1.38 \log{Z/Z_{\odot}} + 29.17$, which is applicable for $5.4 \leq \log{L_{\rm bol}/L_{\odot}} \leq6.1$ and $14 \leq T_{\rm eff}/{\rm kK} \leq 32$.}
   {We find a significant dependence of the wind momentum on the metallicity, which is largely due to the mass-loss rates. We do not find any evidence of a discontinuity in either the mass-loss rate or the ratio of the terminal wind velocity to the escape velocity, $\varv_{\infty}/\varv_{\rm esc}$, between $25$ and $21$~kK, which could be attributed to the bi-stability jump, although when taking into account the effect of luminosity in the transformed mass-loss rate, the behaviour appears to be different. Stellar parameters are consistent across different methods and radiative transfer codes, whereas mass-loss rates differ significantly with our values being generally lower. We find a discrepancy between the evolutionary and spectroscopic masses in $40\%$ of our sample, with the evolutionary mass usually being systematically higher. The mass-loss rates of blue supergiants are far too low to strip the stellar envelope and the subsequent formation of classical Wolf-Rayet (WR) stars, leading to the conclusion that luminous blue variable eruptions or binary interactions are necessary to explain characteristics of the WR population in the SMC.}

   \keywords{stars: massive, stars: early-type, stars: mass-loss, supergiants, stars: atmospheres, stars: winds, outflows}

   \maketitle

\section{Introduction}
Massive stars $(M > 8 M_{\odot})$ possess strong outflows of material, which form due to the immense radiative pressure overtaking the force of gravity at the outer layers of the star. This allows the star's atmosphere to expand beyond the boundaries of the photosphere. The properties of these outflows are strongly correlated to the ratio of luminosity to mass of the star \citep{eddington1926}. To overcome gravity, the acceleration due to free electrons has to be combined with so-called `line-driving', i.e. momentum transfer from radiation absorbed (and re-emitted) in spectral lines \citep{lucy&solomon1970, CAK}. This means that the momentum driving the winds of massive stars is transferred via spectral lines. This leads to the conclusion that the chemical content of the star, or the metallicity ($Z$), plays an important role in determining the properties of the winds \citep{abbott1982}. This is particularly true for metals, and especially iron-like species, which account for the majority of line driving. This is due to their complex atomic structure and the large number of line transitions in their ions. 

The theory of line-driven wind and its implications for the $Z$ dependence of wind properties have been studied in the literature. It is taken into account in the various numerical mass-loss recipes \citep[e.g.][]{vink1999, vink2001, krticka2021}. The $Z$ dependence of mass-loss and wind velocity has also been well established observationally \citep[e.g.][]{kudritzki1987, prinj&crowther1998, mokiem2007, ramachandran2019, hawcroft2024}. Nevertheless, there is a persistent discrepancy between the $Z$ dependence that is predicted in theory and what is empirically derived from observations of blue supergiants in low-$Z$ environments \citep{krticka2024}.

The effects of $Z$ extend beyond the academic interest in the wind properties of massive stars. Mass loss in massive stars is one of the factors that determines the evolutionary path of the star, including whether it is fated to explode in a type of core-collapse supernova (ccSNe II/Ib/Ic) and leave behind a black hole (BH) or a neutron star (NS) remnant \citep{smartt2009}, or whether it would experience direct collapse into a BH in a failed-nova scenario \citep[for a review of the effects of mass-loss on the evolution of massive stars, see e.g.][]{smith2014}.

The winds of massive stars also play an important role in the chemical evolution of their host galaxies by enriching the interstellar medium (ISM) with heavy elements that are synthesised in their interiors and are transported to the surface of the star via efficient internal mixing processes \citep{langer2012}. The mass ejected via stellar winds propagates through the ISM at supersonic speeds, leading to strong mechanical feedback to the surroundings of massive stars. This, in addition to the enrichment of the ISM in heavy elements, leads to an increase in its opacity, affecting the star formation rate in the host galaxy \citep[for a general review on massive star feedback see e.g.][]{geen2023}. 

Due to the abundant metal lines in B-type supergiants, they have been used to constrain heavy metal content in extragalactic environments \citep{urbaneja2005a, urbaneja2005b,trundle2004, trundle2005, bresolin2022}. They have also been successfully used as distance indicators in external galaxies due to their high visual brightness \citep{kudritzki2003, kudritzi2024}. 

This study was made possible by the advent of the Hubble Ultra-violet Legacy Library of Young Stars as Essential Standards \citep[HST ULLYSES,][]{roman-duval2025}. The HST ULLYSES programme dedicated 500 orbits to obtain high-resolution UV spectra of OB stars in low-Z environments, mainly the Large Magellanic Cloud (LMC) and Small Magellanic Cloud (SMC). The UV range is critical for determining the properties of the wind, but it lacks information about the underlying photosphere. For a fuller picture of the stellar and wind properties, we made use of the XShooting/ULLYSES optical spectroscopic legacy library \citep[XShootU,][]{Sana2024}.

Recent efforts have been made by \citet{marcolino2022}, \citet{backs2024}, and \citet{pauli2025} to constrain the $Z$ dependence of wind properties and to produce empirical recipes for mass loss and wind momentum. These recipes, although very insightful, suffer from different issues. In particular, these recipes are produced using assumptions inferred from numerical recipes \citep{leitherer1992, vink2001, krticka2021, bjorklund2023}. Furthermore, luminosity classes are not taken into account in these recipes. Lastly, empirical recipes usually use results from the literature, which use different atmosphere codes and various datasets, some UV and optical, and some only optical, subjecting the derived parameters to large variance \citep{sander2024}.

The present study may be viewed as the second in a series of papers, following \citet[][henceforth Paper~XIII]{alkousa2025}, in which a sample of LMC late-O and B supergiants was analysed. In the present study, we expand our analysis of OB stars to the SMC. We aim to quantify the effect of $Z$ on the properties of the wind. We do not aim to produce a full empirical recipe dependent on $Z$ of mass loss, as doing this without forcing any assumptions from numerical simulations would require another $Z$ environment, such as the Milky Way (MW), which we aim to undertake in a future study. 

The XShootU and ULLYSES datasets have been used in recent studies of OB supergiants in the SMC, employing various radiative transfer codes and fitting techniques for different purposes. We compare our derived photospheric and wind parameters to those obtained by \citet{Bestenlehner2025}, \citet{backs2024}, and \citet{bernini2024} in Appendix~\ref{app:comp}.

In Section~\ref{sec:observations}, we present the dataset that was used in this study, our sample, and our selection criteria. In Section~\ref{sec:methods}, we give a summary of our methodology, which was presented in Paper~XIII in thorough detail. In Section~\ref{sec:results}, we introduce the results of our analysis. In Section~\ref{sec:discussion}, we discuss the implications of our results, including the effects of $Z$ and the bi-stability jump. 
\section{Observations}
\label{sec:observations}
\label{observations}
   \begin{table*}
        \caption{List of the stars included in our analysis and the respective wavelength coverage.}
        \label{table:data}    
        \centering            
        \small
        \begin{tabular}{c c c c c c c c}        
        
        \hline
        Star           &Alias              &SpT     &$(900-1160)$   &$(1150-1700)$   &$(1700-2370)$ &Aux. Opt.\\
                       &         &                  &FUSE           &HST             &HST           &Magellan\\
        \hline 
         AzV\,469   &Sk\,148               &O9 Iab(f)&LWRS          &STIS E140M      &-             &-    \\ 
         AzV\,372   &Sk\,116               &O9.2 Iab&LWRS           &STIS E140M      &STIS E230M    &-    \\
         AzV\,456   &Sk\,143               &O9.5 Iab&LWRS           &COS G130M+G160M &STIS E230M    &-    \\ 
         AzV\,327   &R\,28, BLOeM 1-066    &O9.7 Ib &LWRS           &STIS E140M      &-             &MIKE \\ 
         AzV\,235   &R\,17, Sk\,82, BLOeM 7-064    &B0 Ia   &LWRS           &STIS E140M      &STIS E230M    &-    \\ 
         AzV\,215   &Sk\,76                &B0 Ia   &LWRS           &STIS E140M      &STIS E230M    &-    \\ 
         AzV\,104   &-                     &B0.5 Ia &LWRS           &STIS E140M      &STIS E230M    &MIKE \\ 
         AzV\,410   &-                     &B0.7 Iab&-           &STIS E140M      &-             &-    \\ 
         AzV\,242   &R\,18, Sk\,85, BLOeM 4-078    &B0.7 Ia &LWRS           &STIS E140M      &STIS E230M    &-    \\ 
         AzV\,96    &Sk\,46, BLOeM 8-008   &B1 Iab  &LWRS           &STIS E140M      &STIS E230M    &MIKE \\
         AzV\,264   &Sk\,94, BLOeM 1-009   &B1 Ia   &LWRS           &STIS E140M      &STIS E230M    &MIKE \\
         AzV\,175   &Sk\,64                &B1.5 Ib &LWRS           &COS G130M+G160M &-             &MIKE \\
         Sk\,191    &-                     &B1.5 Ia &LWRS           &STIS E140M      &STIS E230M    &-    \\ 
         AzV\,18    &Sk\,13                &B2 Ia   &LWRS           &STIS E140M      &STIS E230M    &MIKE \\
         NGC330\,ELS\,4 &Rob\,B37          &B2.5 Ib &-           &STIS E140M      &STIS E230M    &MIKE \\ 
         AzV\,187   &Sk\,68                &B2.5 Ia &LWRS           &COS G130M+G160M &STIS E230M    &MIKE \\ 
         AzV\,22    &2dFS 5015             &B3 Ia   &-              &STIS E140M      &STIS E230M    &-    \\ 
         AzV\,393   &R\,39, Sk\,124                 &B3 Ia$^+$&LWRS          &COS G130M+G160M &STIS E230M    &MIKE \\ 
         AzV\,343   &Sk\,111               &B8 Iab  &-           &COS G130M+G160M &STIS E230M    &MIKE \\ 
         AzV\,324   &BLOeM 1-062           &B8 Ib   &-           &COS G130M+G160M &STIS E230M    &-    \\ 
        \noalign{\smallskip}
            \hline
        \end{tabular}
        \tablefoot{The spectral types are based on XShootU observations \citep{Bestenlehner2025}.}
    \end{table*}
    
In Table~\ref{table:data}, we present the stars in our sample, along with their classification, which was adopted from \citet{Bestenlehner2025}. In this table, we also include the telescope, instrument, and wavelength coverage available for each star. 

This sample was selected in such a way that it would cover a spectral range similar to that of Paper~XIII. We cover a wide range of spectral classes, from O9 to B8, which includes the region of the theorised bi-stability jump. We attempted to include more than one star per spectral class, ideally with different luminosity classifications (Ia$^{+}$, Ia, Iab, Ib). This allowed us to explore the $Z$ dependence of wind properties while avoiding selection bias, which we discussed in Paper~XIII when contrasting our results for the LMC sample with the results of \citet{bernini2024} in the SMC, whose sample was made up almost exclusively of Ia supergiants. We have ten stars in common with \citet{bernini2024}, who include eight stars that are not covered in our study, while we include ten stars not covered in \citet{bernini2024}. That makes the two studies complementary, although with slightly varying modelling techniques (X-rays and micro-turbulence), and an excellent opportunity to compare the results in our overlapping samples.   

A subset of our sample has been included in the BLOeM survey \citep{shenar2024}. \citet{britavskiy2025} did not find evidence of significant radial velocity shifts across nine epochs of this subset, indicating that these stars are apparently single. They report that AzV\,242 (BLOeM 4-078) and AzV\,96 (BLOeM 8-008) show intrinsic line profile variability, with the possibility that AzV\,242 is an SB1 spectroscopic binary. 

\subsection{ULLYSES}

The ULLYSES programme \citep{roman-duval2025} made use of two instruments aboard HST ; namely, the Cosmic Origins Spectrograph \citep[COS, ][]{COS} in the G130M/1291, G160M/1611, and G185M/1953 wavelength settings, achieving a minimum spectral resolving power of $R\approx12,000$, $R\approx13,000$, and $R\approx16,000$, respectively, and the Space Telescope Imaging Spectrograph \citep[STIS, ][]{STIS} E140M/1425 $(R\approx46,000)$, and E230M/1978 $(R\approx30,000)$ gratings in the far- and near-UV. New COS and STIS spectra were obtained for a subset of our targets between July 2020 and Sept 2022, and were combined with archival HST spectra drawn from GO 7437 (P.I. Lennon), GO 9116 (P.I. Lennon), GO 12581 (P.I. Roman-Duval), GO 13778 (P.I. Jenkins), and GO 15837 (P.I. Oskinova) obtained between Oct 2001 and Jun 2020. A subset of HST spectra have previously been presented by \citet{Walborn2000} and \citet{Evans2004}.

Archival spectra from the Far Ultra-violet Spectroscopic Explorer \citep[ FUSE,][]{FUSE} cover the wavelength range of $\approx 900-1160~\AA$ and uniformly involve the LWRS ($30'' \times 30''$) aperture for our sample, providing $R\approx 15,000$. These are drawn from various Principal Investigator Team and Guest Investigator programmes obtained between May 2000 and July 2003. FUSE spectroscopy of a subset of the current sample has previously been presented by \citet{Walborn2002}. We note that AzV\,327 lies in a crowded region, such that other sources, including [M2002]\,SMC\,55495, contribute to the FUSE spectrum \citep{danforth2002}.

\subsection{XShootU} 

ULLYSES targets were observed using the X-shooter instrument \citep{vernet2011}, which is mounted on the Very Large Telescope (VLT). This slit-fed spectrograph ($11 \arcsec$ slit length) provides simultaneous coverage of the wavelength region between $3000$-$10200~{\rm \AA}$, across two arms: UVB ($3000 \le \lambda \le 5600~{\rm \AA}$) and VIS ($5600 \le \lambda \le 10200~{\rm \AA}$). The observations were carried out between October and December 2020 with a slit width $0.8\arcsec$ for the UBV arm, achieving $R\approx6700$, and $0.7\arcsec$ for the VIS arm, with $R \approx 11400$ \citep{xshootu1}. All science-ready optical spectra used in this study were combined, flux-calibrated, corrected for telluric contamination, and normalised by \citet[DR1]{Sana2024}, except for archival X-shooter spectroscopy of AzV\,187 from November 2013, which we normalised using a higher-order polynomial fit of points selected around regions of clean continuum. 

\subsection{Photometry} 
The photometric magnitudes used in this study were taken from the compilation in \citet{xshootu1} to derive the bolometric luminosities by fitting the SED spectral energy distribution. The $U B V$ photometry was taken from \citet{ardeberg1977}, \citet{azzopardi1975}, \citet{ardeberg1980}, and \citet{massey2002}, and the $J Ks$ photometry was drawn from the VISTA near-infrared $Y J K_{S}$ survey of the Magellanic System \citep[VMC,][]{VMC}, with $H$-band photometry taken from the Two Micron All Sky Survey \citep[2MASS,][]{cutri2003, skrutskie2006, cutri2012}.

\subsection{MIKE spectroscopy} 
High-resolution optical spectroscopy used to extract the rotational properties of slow rotators was collected between December 2021 and December 2022 using the Magellan Inamori Kyocera Echelle (MIKE) spectrograph mounted on the Magellan Clay 6.5~m Telescope, covering the wavelengths $3350$ to $5000~\AA$ in the blue arm and $4900$ to $9500~\AA$ in the red arm, achieving $R\approx 35000-40000$ \citep{mike}.

\section{Methods}
\label{sec:methods}
In this section, we present a summary of the modelling and fitting techniques used in our analysis.  In the present study, we employ the same strategies as in Paper~XIII. Therefore, for a comprehensive and detailed description of our methodology, we refer the reader to Paper~XIII. 

\subsection{Model atmosphere and grid}
We employed the 1D radiative transfer code \textsc{CMFGEN} \citep{Hillier1990, hillier1998}, which solves the radiative transfer equations in spherical symmetry in non-local thermodynamic equilibrium (non-LTE), taking into account the effects of mass loss under the assumption of purely radial outflows. \textsc{CMFGEN} also accounts for the effects of extreme UV blanketing caused by millions of lines of iron-like species and other elements on populations and the ionisation structure in the wind. This is implemented using super-levels \citep{anderson1985, anderson1989}, which allow for the bundling of levels with similar energies and properties into a single or a super-level. This significantly reduces the number of statistical equilibrium equations that must be explicitly solved.

\textsc{CMFGEN} does not calculate the velocity stratification in the wind consistently with radiative acceleration, but rather superimposes a velocity law with a simple $\beta$ parametrisation. Additionally, wind inhomogeneities are considered under the `optically thin clumping’ approximation, or micro-clumping, assuming that these optically thin clumps are smaller than the mean free path of the photons \citep{hillier1996} and that the medium between the clumps is completely void of matter \citep{hillier1997, HillierandMiller1999}.
  
\begin{table}
        \caption{CMFGEN SMC-metallicity grid parameters.}
        \label{table:grid}    
        \centering            
        \small
        \begin{tabular}{c c c}        
        \hline
        \multicolumn{3}{c}{Iterated parameters} \\
        \hline
        parameter                                   &Values             &Step\\
        \hline\\[0.15 pt]
        $\log{(T_{\rm eff}/{\rm K})}$               &4.150 $\cdots$ 4.600     &0.025\\
        $\log{(g/{\rm cm\,s^{-2}})}$                &1.7 $\cdots$ 3.9         &0.2\\
        $\log{(\dot{M}/M_{\odot}\,{\rm yr}^{-1})}$  &--7.0 $\cdots$ --5.5       &0.3\\ 
        \hline
        \multicolumn{3}{c}{Fixed parameters} \\
        \hline
        parameter                                   &\multicolumn{2}{c}{Value}\\
        \hline\\[0.15 pt]
        $\log{(L/L_{\odot})}$                       &\multicolumn{2}{c}{5.8}   \\
        $\beta$                                     &\multicolumn{2}{c}{1.0}   \\ 
        $f_{\rm vol,\infty}$                               &\multicolumn{2}{c}{0.1}   \\
        $\epsilon_{\rm C}$                          &\multicolumn{2}{c}{7.42}  \\              
        $\epsilon_{\rm N}$                          &\multicolumn{2}{c}{6.66}  \\
        $\epsilon_{\rm O}$                          &\multicolumn{2}{c}{8.05}  \\
        $\epsilon_{\rm Ne}$                         &\multicolumn{2}{c}{7.23}  \\
        $\epsilon_{\rm Mg}$                         &\multicolumn{2}{c}{6.78}  \\
        $\epsilon_{\rm Al}$                         &\multicolumn{2}{c}{5.59}  \\
        $\epsilon_{\rm Si}$                         &\multicolumn{2}{c}{6.72}  \\
        $\epsilon_{\rm P}$                          &\multicolumn{2}{c}{5.11}  \\
        $\epsilon_{\rm S}$                          &\multicolumn{2}{c}{6.36}  \\
        $\epsilon_{\rm Ca}$                         &\multicolumn{2}{c}{5.61}  \\
        $\epsilon_{\rm Fe}$                         &\multicolumn{2}{c}{6.80}  \\
        $\epsilon_{\rm Ni}$                         &\multicolumn{2}{c}{5.37}  \\
        \hline
        \multicolumn{3}{c}{Dependent parameters} \\
        \hline\\[0.15 pt]
        $\varv_{\infty}(T_{\rm eff})$               &\multicolumn{2}{c}{\citet{hawcroft2024} recipe}\\
        \hline
        \end{tabular}
        \tablefoot{The values for abundance were adopted from \citet{xshootu1}, who obtained a mean baseline SMC from multiple studies that derived the elemental abundances using different methods.}
    \end{table}

In Table~\ref{table:grid}, we present the properties of our grid. The effective temperature, $T_{\rm eff}$, surface gravities, $\log{(g/{\rm cm\,s^{-2}})}$, and mass-loss rates, $\log{(\dot{M}/M_{\odot}\,{\rm yr}^{-1})}$, were refined starting from the preferred grid model. Other parameters, such as elemental abundances ($\epsilon_{\rm X} = \log{X/H}+12$), wind acceleration parameter, $\beta$, volume filling factor, $f_{\rm vol,\infty}$, and terminal wind velocity, $\varv_{\infty}$, were fixed in the grid, but were later fine-tuned for each star. Stars that have $T_{\rm eff}$ or $\log{g}$ values outside of the range covered by the grid were handled by building miniature grids and using those grids to fine-tune the model parameters.

The atomic model and data used in this study are identical to those shown in Table C.1 in Paper~XIII. We included 14 species and 50 different ions. In cooler models ($<25~{\rm kK} $), we omitted higher ionisation stages and included the lower ionisation stages. This is important for model convergence and for the iron forest in the UV spectra of B stars. 

Also, similarly to Paper~XIII, we excluded X-rays from our fitting procedure to guarantee a homogeneous analysis without variations in the techniques. X-rays, which were included in \citet{bernini2024}, increase the population of high ionisation stages (super-ionisation), affecting the strength of P Cygni resonance lines in the UV, such as $\ion{N}{V}~\lambda1238,\,1242$, $\ion{Si}{IV}\lambda\lambda1394,\,1403$, and $\ion{C}{IV}~\lambda\lambda1548,\,1551$. The super-ionisation comes at the expense of depopulating the lower ionisation stages, affecting lines of $\ion{Al}{III}$ and $\ion{C}{II}$.

In our grid, we choose a simple modified $\beta$ velocity law that was introduced in \citet{castor&lamers1979}:
\begin{equation}
  \label{eq:vel_law} 
    \varv(r) = \varv_{0}+(\varv_{\infty}-\varv_{0})\left(1-\frac{R_{\ast}}{r}\right)^{\beta},
\end{equation}
where $\varv_{\infty}$ is the terminal wind velocity, $\varv_{0}$ is the connection velocity, which is estimated as two-thirds the speed of sound $\approx10~{\rm km\,s^{-1}}$, and $R_{*}$ is the radius of the star, which is defined at the Rosseland optical depth $\tau = 100$. The velocity adopted for each point on the grid was obtained via the $T_{\rm eff}$ and $Z$-dependent empirical recipe from \citet{hawcroft2024}:
\begin{equation}
  \label{eq:vinf_callum} 
    \varv_{\infty}~({\rm km\,s^{-1}}) = \left[0.092(\pm 0.003)T_{\rm eff}~({\rm K})-1040(\pm 100)\right]Z/Z_{\odot}^{(0.22\pm 0.03)},
\end{equation}
which was derived from Sobolev with exact integration (SEI) modelling of the P Cygni resonance doublet $\ion{C}{IV}~\lambda\lambda1548-1551$ to measure the terminal wind velocity. We adopted $Z= 0.2~Z_{\odot}$ in Equation~\ref{eq:vinf_callum} \citep{xshootu1}. 

To account for wind inhomogeneity, we adopted an exponential clumping law of the form
\begin{equation}
  \label{eq:clumping} 
    f(r) = f_{\rm vol,\infty} + (1-f_{\rm vol,\infty})\exp{\left(-\frac{\varv (r)}{\varv_{\rm cl}}\right)},
\end{equation}
where $f_{\rm vol,\infty}$ is the terminal volume-filling factor and $\varv_{\rm cl}$ is the onset clumping velocity. We adopted the values $f_{\rm vol,\infty} = 0.1$ and $\beta = 1$ in the grid, both of which were later modified in the fine-tuning process to obtain satisfactory fits for the H$\alpha$ and UV P Cygni lines. 

The grid spans a range for the transformed radius, $\log{R_t}$ \citep{schmutz1989}, of between $2.1$ and $4.0$ on a logarithmic scale, where $R_t = R_{\ast}(\varv_{\infty}\,\dot{M}^{-1}\, f_{\rm vol}^{0.5}\,10^{-4}~M_{\odot}{\rm yr}^{-1}/2500~{\rm km\,s^{-1}})^{2/3}$, with smaller values relating to denser winds. This $\log{R_t}$ range corresponds to a range of optical depth-invariant wind-strength parameters, $\log{Q} = \log{\dot{M}/(R_{\ast} \varv_{\infty})^{3/2}}$ \citep{puls1996}, of $-15.0$ to $-12.5$, where larger values of $\log{Q}$ correspond to denser winds. We employed $R_t$ to scale $\dot{M}$ to the derived bolometric luminosity of the star.

\subsection{Fitting procedure and diagnostics}
We first used the results of \citet{Bestenlehner2025} to pinpoint the closest fitting model on the grid in $\log{T_{\rm eff}}$-$\log{g}$-$\log{\dot{M}}$ parameter space, after which we start the fine-tuning procedure. The analysis in \citet{Bestenlehner2025} is based on the model de-idealisation pipeline, which applies a minimisation and interpolation routine using a large grid of \textsc{FASTWIND} models \citep{Puls2005, rivero2011}. This allows for the estimation of stellar parameters and the mass-loss rates and model uncertainties using only optical spectroscopy \citep{bestenlehner2024}. 

Once we selected the nearest grid model, we started by fine-tuning the values of $T_{\rm eff}$, $\log{g}$, and the helium abundance iteratively until a satisfactory fit was achieved. Then we iteratively fine-tuned $\log{\dot{M}}$, $\beta$, $f_{\rm vol,\infty}$, and $\varv_{\infty}$ to fit H$\alpha$ and the UV P Cygni lines. Finally, the CNO mass fractions were fine-tuned for each star. 

Here, we give a summary of the fitting procedure. The approach for determining the goodness of the fit was `$\chi$-by-eye'. For an extensive description of the fitting techniques, diagnostics, and uncertainty determination, we refer the reader to Section 3 of Paper~XIII. 

\paragraph{Effective temperature, $T_{\rm eff}$} We obtained the effective temperature from the principle of ionisation equilibrium. In O supergiants the primary $T_{\rm eff}$ diagnostics are $\ion{He}{I}~\lambda4471$--$\ion{He}{II}~\lambda4542$ \citep{martins2011}. For early and mid-B supergiants, we used ratios of adjacent ions of silicon as primary diagnostics \citep{mcerlean1999}; namely, $\ion{Si}{IV}~\lambda\lambda4088-4116$ versus $\ion{Si}{III}~\lambda\lambda\lambda4553-4568-4575$ for B0 to B2 stars, and $\ion{Si}{III}~\lambda\lambda\lambda4553-4568-4575$ versus $\ion{Si}{II}~\lambda\lambda4128-4131$ for B2.5-B3 stars. Note that in early B stars the $\ion{Si}{IV}~\lambda4088.96$ line is blended with $\ion{O}{II}~\lambda4089.29$ \citep{Hardorp&Scholz1970, deburgos2024b}, so we ensured that this transition is accounted for in the adopted atomic model. For B8 supergiants, we used the ratio of $\ion{He}{I}\lambda4471$ to $\ion{Mg}{II}~\lambda4481$ to determine $T_{\rm eff}$. 

\paragraph{Surface gravity, $\log{g}$} We determined the surface gravity by fitting the wings of Balmer lines, which are very sensitive to Stark broadening. The primary $\log{g}$ diagnostics for all stars in our sample are H$\gamma$, H$\eta$, and H$\zeta$. These lines are relatively isolated -- in contrast with H$\delta$ -- and are usually not contaminated by wind effects \citep{lennon1992}. 

\paragraph{Luminosity} To determine the bolometric luminosity, $L_{\rm bol}$, of stars, we used the model fluxes (which were calculated for $\log{L_{\rm model}/L_{\odot}}=5.8$). The reddening law from \citet{gordon2003} was then applied to the model fluxes. We then fitted the relative extinction, $R_{V}$, and other parameters that determine the shape of the UV extinction curve \citep{FM1990}. Then the intrinsic colours were obtained ($B^{m}$, $V^{m}$, and $Ks^{m}$) by applying filter functions from \textsc{pyphot} \citep{zenodopyphot} to the model SED using the Vega magnitude system photometric zero-point (for more details on \textsc{pyphot}, see Table~3 in Paper~XIII). The colour excess $E(B-V)$ was calculated as $(B - V) - (B^{m} - V^{m})$. The model SED was then scaled to match observations using a factor equal to the ratio of $Ks$ fluxes $F_{Ks}^{m}/F_{Ks}$. The bolometric correction, $BC_{V}$, was calculated as $-2.5\log{L_{\rm model}} + 4.74 - V^{m}$. The final $L_{\rm bol}$ was calculated as  $(DM + A_{V} - BC^{m}_{V} - m_{V} + 4.74)/2.5$, where $A_{V} = R_{V}\cdot E(B-V)$, and assuming a distance modulus of $DM = 18.96$ (corresponding to a distance $d=62~{\rm kpc}$) for the SMC \citep{scowcroft2016}. In other relevant XShootU studies, a value of $18.98$ for the distance modulus was adopted from \citet{graczyk2020}.

\paragraph{Helium mass fraction} We obtained the helium abundance by fitting $\ion{He}{I}$ and $\ion{He}{II}$ lines relative to hydrogen lines. The diagnostic lines used across the sample are $\ion{He}{I}~\lambda4026$, $\ion{He}{I}~\lambda4471$, and $\ion{He}{I}~\lambda4922$ for O and B supergiants, and $\ion{He}{II}~\lambda4542$ and $\ion{He}{II}~\lambda5411$ for O supergiants. We also used $\ion{He}{I}~\lambda6678$, $\ion{He}{I}~\lambda7065$, and $\ion{He}{I}~\lambda7281$ as a sanity check.

\paragraph{Wind density parameters} The wind density, $\rho$, is defined by $\varv_{\infty}$, $\log{\dot{M}}$, $f_{\rm vol,\infty}$, $\beta$, and $\varv_{\rm cl}$. The primary optical diagnostic line for $\log{\dot{M}}$ and $\beta$ is H$\alpha$ $(\propto \rho^{2})$. The UV unsaturated P Cygni profiles provide valuable information on wind inhomogeneities \citep{massa2008, searle2008, puls2008}. The UV lines that we used to constrain $\log{\dot{M}}$ and $f_{\rm vol,\infty}$ are $\ion{S}{IV}\lambda\lambda1063-73$, $\ion{C}{III}\lambda1176$, $\ion{Si}{IV}\lambda\lambda1394-1403$, $\ion{C}{IV}~\lambda\lambda1548-51$, and $\ion{Al}{III}\lambda\lambda1856-62$. $\varv_{\rm cl}$ was adopted as two times the speed of sound of the model. $\varv_{\infty}$ was obtained from either (i) direct measurements of the black velocity, $\varv_{\rm black}$ (the velocity measured at the bluest extent of fully saturated P Cygni absorption \citep{prinja1990}), or (ii) as a fraction of $\varv_{\rm edge}$ (the velocity measured at the point where the blue trough of the P Cygni profile intersects the local continuum \citep{abbot1985}). We present the results of the velocity measurement in Table.~\ref{table_app_2}. 

\paragraph{Line broadening parameters} Absorption lines are broadened by the effect of rotation \citep{slettebak1956}, photospheric macro-turbulence, $\varv_{\rm mac}$ \citep{contiandebbets1977, simon-diaz2010, simon-diaz2014}, and micro-turbulence, $\varv_{\rm mic}$ \citep{mcerlean1998}. We fixed $\varv_{\rm mac}=20~{\rm km\,s^{-1}}$ and $\varv_{\rm mic}=10~{\rm km\,s^{-1}}$, and only determined the projected rotational velocity, $\varv_{\rm rot}\sin{i}$, using metal lines. For O stars, our primary diagnostic is $\ion{O}{III}~\lambda 5592$. For early B stars, we used $\ion{C}{III}~\lambda 4267$ and $\ion{Si}{iii}~\lambda 4553$. For mid- to late-B stars, we fitted the $\ion{N}{II}~\lambda 3995$ line and, as a sanity check, we used $\ion{Si}{II}~\lambda6347$, the $\ion{Mg}{II}~\lambda 4481$ doublet, and $\ion{C}{II}~\lambda 4267$. 

\paragraph{CNO abundances, $\epsilon_{\rm X}$} The abundances of carbon, nitrogen, and oxygen were constrained by fitting multiple optical absorption lines of each element. The selection of lines depends on the ion that dominates the photosphere in each spectral type. In Table~\ref{table:CNO}, we present our choice of diagnostic lines. The CNO abundances (among other metals) are arguably the most sensitive parameters to changes in photospheric micro-turbulent velocity, $\varv_{\rm mic}$ \citep{urbaneja2005b}. Fixing $\varv_{\rm mic}$ in our analysis reduces the precision of our derived CNO abundances. This is due to the effect that changing the photospheric $\varv_{\rm mic}$ has on the opacity, and therefore on the line strength of different components of multiplets \citep{mcerlean1998}.

\section{Results}
\label{sec:results}
In this section, we present an overview of the results of our analysis. We compare our results with those from previous studies of SMC blue supergiants. In addition, we compare the derived wind parameters with predictions from numerical recipes. In Table~\ref{table:2}, we present the derived physical parameters and the inferred evolutionary masses and ages. In Appendix~\ref{app:comp}, we compare our derived stellar and wind parameters to values from the literature. Comments on the spectral fitting quality are provided for each star individually in Appendix~\href{https://doi.org/10.5281/zenodo.18416382}{G}. The SED fits, individual line fits, and overall spectral fits can be found in Appendix~\href{https://doi.org/10.5281/zenodo.18416382}{H}, \href{https://doi.org/10.5281/zenodo.18416382}{I}, and \href{https://doi.org/10.5281/zenodo.18416382}{J}, respectively.

\subsection{Hertzsprung-Russell diagram}
\label{sec:HRD}
\begin{figure}
    \centering
     \includegraphics[width=\hsize]{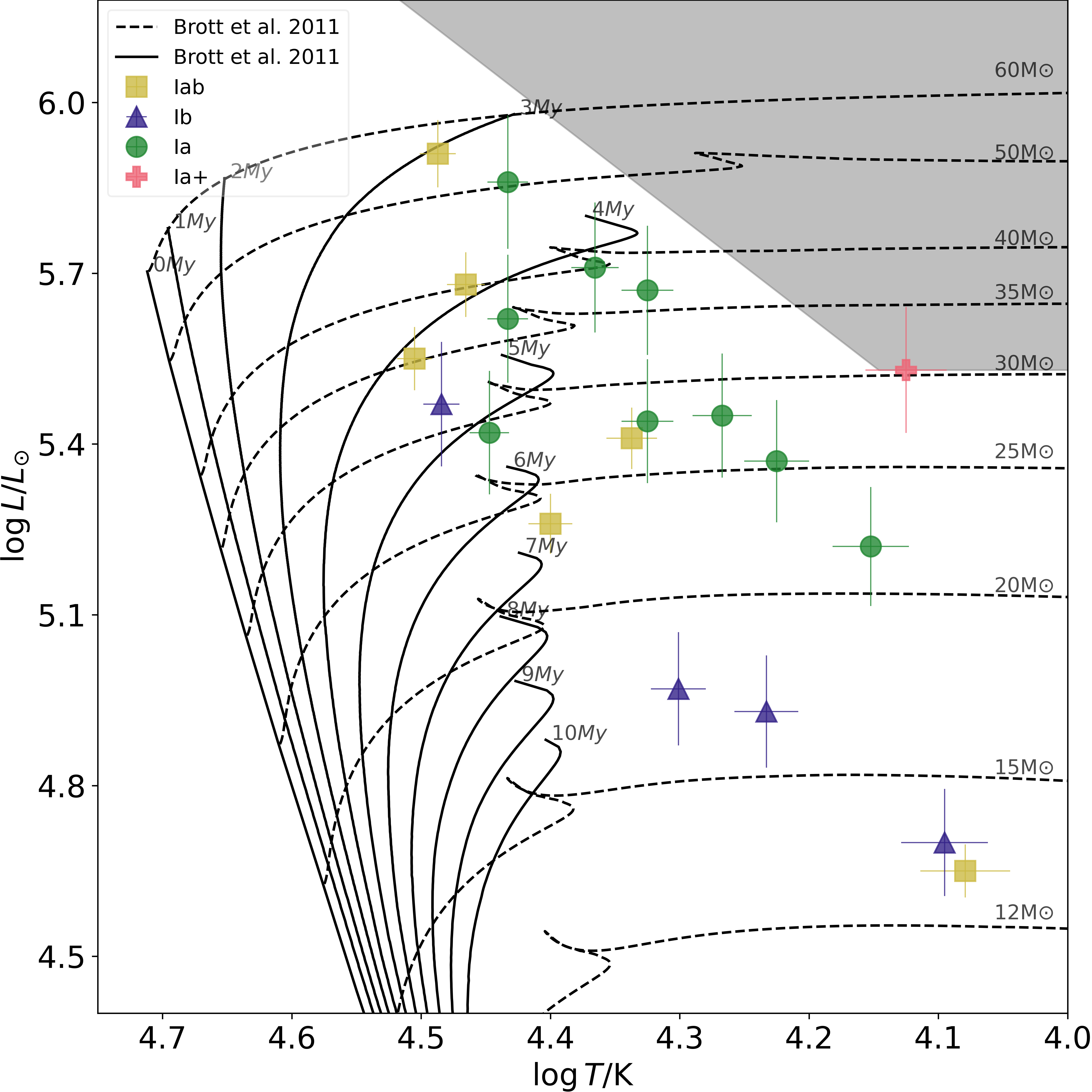}
         \caption{Hertzsprung-Russell diagram for our sample. Overlaid in solid black lines are non-rotating SMC isochrones for different ages ($\approx0{\rm -}10$ Myr). Dashed black lines are the SMC rotating evolutionary tracks for stellar masses in the range of $\approx10{\rm -}60~M_{\odot}$ with a rotational velocity of $110~{\rm km\,s^{-1}}$. Both the isochrones and evolutionary tracks are adopted from \citet{brott2011}. The shaded area is defined by the HD limit \citep{humphreys&davidson1979, smith2004, davies2018}.}
         \label{HRD}
    \end{figure}
    
Fig.~\ref{HRD} shows the Hertzsprung-Russell diagram (HRD) for our sample, superimposed on SMC isochrones (solid black lines) and evolutionary models (dashed black lines) \citep{brott2011}. Our sample spans a wide range of $\log{(L_{\rm bol}/L_{\odot})}$, from $4.65$ (AzV\,\,343) to $5.91$ (AzV\,\,456), and a $T_{\rm eff}$ range of $12.0~{\rm kK}$ (AzV\,343) to $32.0~{\rm kK}$ (AzV\,469). The shaded area in Fig.~\ref{HRD} is defined by the upper limit on the observed luminosities of SMC cool supergiants, also known as the Humphreys-Davidson (HD) limit \citep{humphreys&davidson1979, smith2004, davies2018}. 

As a sanity check of $T_{\rm eff}$, we compared the predicted Balmer jump strengths to observations \citep{kudritzki2008, urbaneja2017}.  We find that the strength of the predicted Balmer jumps -- the $T_{\rm eff}$ of which was obtained from the ionisation balance -- agrees well with observations within a few hundred Kelvin.

The O supergiants in our sample occupy the region $\log{T_{\rm eff}/{\rm kK}}>4.48$ (or $T_{\rm eff}> 30$~kK), and according to the isochrones of \citet{brott2011} are still in their main-sequence (MS) phase. In contrast, the distribution of the cooler B supergiants ($\log{T_{\rm eff}/{\rm kK}}<4.3$) on the HRD hints at them being post-MS objects, according to \citet{brott2011} single-star evolutionary models. The evolutionary status of hot BSGs ($4.3<\log{T_{\rm eff}/{\rm kK}}<4.48$) remains ambiguous, as they are in close proximity to the terminal-age main sequence (TAMS).

\subsection{Stellar masses}
\label{sec:mass}
\begin{figure}
    \centering
     \includegraphics[width=\hsize]{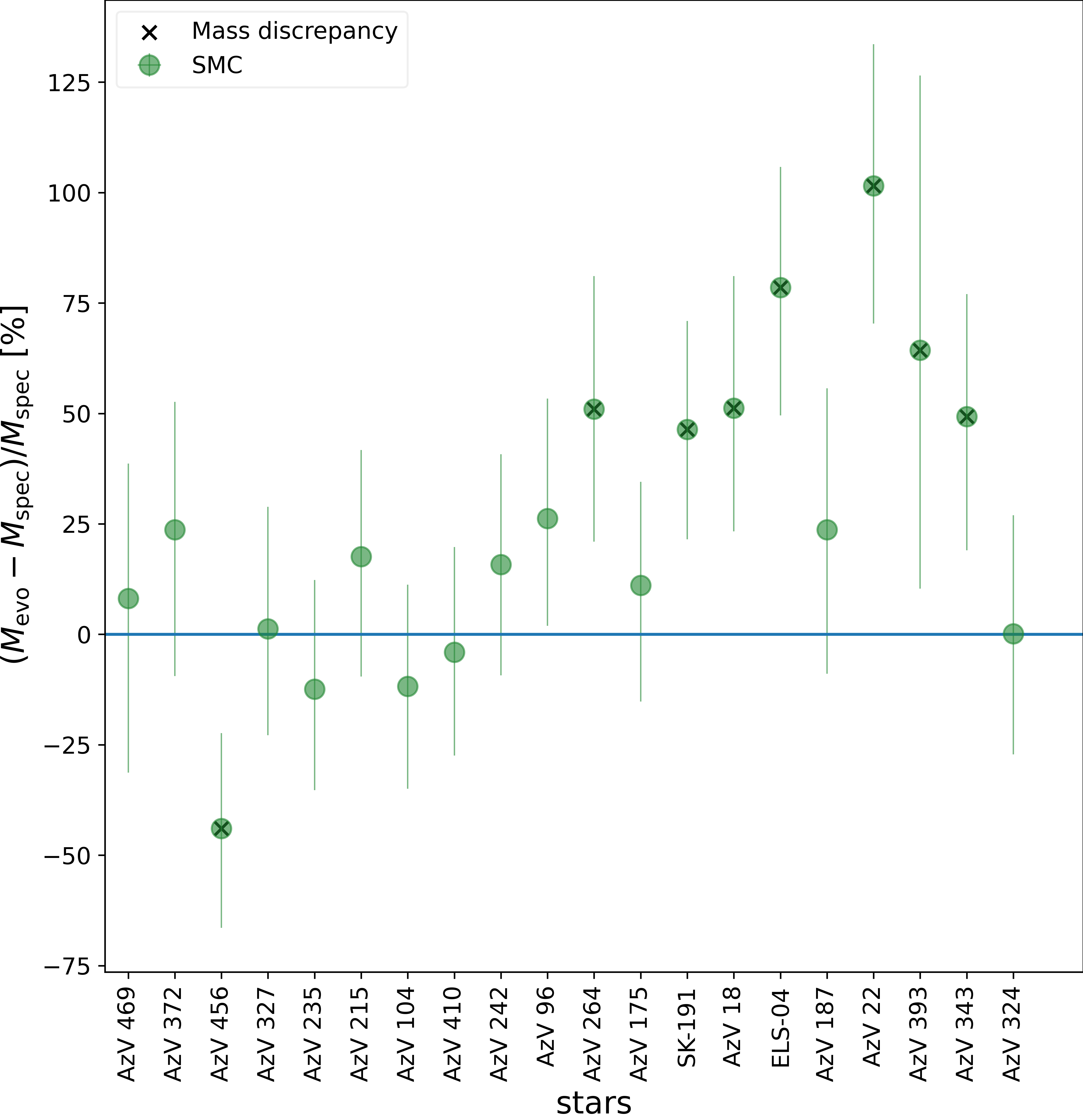}
         \caption{Relative (percentage) difference of the evolutionary and spectroscopic masses $(M_{\rm evo}-M_{\rm spec})/M_{\rm spec}$. $M_{\rm evo}$ was obtained using a Bayesian inference method applied to SMC evolutionary models of \citet{brott2011}. The diagonal black crosses indicate the presence of a mass discrepancy.}
         \label{Mass_comp}
    \end{figure}
    
Spectroscopic masses, $M_{\rm spec}$,  presented in Table.~\ref{table:2}, were calculated via $g = G\,M/R_{*}^{2}$. Since our stars are generally slow rotators, the centrifugal-force corrected gravity, $\log{(g_c/{\rm cm\,s^{-2}})} = \log{(g + \varv_{\rm rot}\sin{i}^{2}/R_{*})}$ \citep{herrero1992}, is very similar to $\log{g}$. By way of example, $\log{g_{c}}-\log{g} = 0.02$~dex for AzV\,372, which has the largest $\varv_{\rm rot} \sin i$ of our sample ($100~{\rm km\,s^{-1}}$). We also present evolutionary masses, $M_{\rm evo}$, and ages, which were obtained using a Bayesian inference method (V.~Bronner et al. in prep) that is similar to {\sc Bonnsai} \citep{bonnsai2014}, applied to \citet{brott2011} rotating single-star evolutionary models for SMC metallicity.

In Fig.~\ref{Mass_comp} we present the relative difference between $M_{\rm spec}$ and $M_{\rm evo}$ for our sample. The error bars representing the uncertainties in Fig.~\ref{Mass_comp} take into account in quadrature the uncertainty of $M_{\rm evo}$, which was obtained from the Bayesian inference method, and the uncertainty of $M_{\rm spec}$, which in relative terms sits between $25\%$ and $35\%$. Generally, $M_{\rm evo}$ is larger than $M_{\rm spec}$ in our sample. We find that $40\%$ of our sample shows a significant discrepancy between $M_{\rm spec}$ and $M_{\rm evo}$. For these stars, $M_{\rm evo}$ is larger than $M_{\rm spec}$, except for AzV\,456, for which we find that $M_{\rm spec}$ ($\approx83\pm12~M_{\odot}$) is significantly larger than $M_{\rm evo}$ ($\approx47^{+6}_{-4}~M_{\odot}$).

\subsection{Wind properties}
\label{subsec:wind_properties}
\begin{table}
        \caption{Derived wind parameters. The $\dagger$ in the superscript indicates an upper limit.}
        \label{table:wind}    
        \def\arraystretch{1.1}  
        \centering      
        \small                               
        \addtolength{\tabcolsep}{-0.35em}
        \begin{tabular}{c c c c c c c c}        
        \hline\smallskip
        Star    &$\log{L_{\rm bol}}$ &$\log{\dot{M}}$ &$\varv_{\infty}$ &$\varv_{\rm esc, 1-\Gamma}$ &$\beta$ &$f_{\rm vol,\infty}$ &$\varv_{\rm cl}$\\
            &$L_{\odot}$ &$M_{\odot}\,{\rm yr}^{-1}$ &${\rm km\,s^{-1}}$ &${\rm km\,s^{-1}}$ & & &${\rm km\,s^{-1}}$\\
        \hline
        AzV\,469        &$5.55$ &$-7.00$ &$1800$ &$628$  &$1.0$  &$0.2$  &$30$\\
        AzV\,372        &$5.68$ &$-6.76$ &$1620$ &$509$  &$1.0$  &$0.1$  &$30$\\
        AzV\,456        &$5.91$ &$-7.57$ &$1313$ &$740$ &$1.0$  &$0.1$  &$30$\\
        AzV\,327        &$5.47$ &$-8.12$ &$1551$ &$658$  &$1.0$  &$0.03$ &$35$\\
        AzV\,235        &$5.86$ &$-6.66$ &$1300$ &$553$  &$1.5$  &$0.05$ &$30$\\
        AzV\,215        &$5.62$ &$-7.17$ &$1585$ &$497$  &$1.7$  &$0.1$  &$35$\\
        AzV\,104        &$5.42$ &$-7.75$ &$996$  &$637$  &$1.0$  &$0.1$  &$35$\\
        AzV\,410        &$5.26$ &$-7.65^{\dagger}$ &$403$  &$564$  &$1.0$  &$0.1$  &$30$\\
        AzV\,242        &$5.71$ &$-7.47$ &$942$  &$424$  &$3.1$  &$0.03$ &$30$\\
        AzV\,96         &$5.41$ &$-7.43$ &$758$  &$409$  &$2.6$  &$0.1$  &$30$\\
        AzV\,264        &$5.44$ &$-7.32$ &$656$  &$363$ &$3.0$  &$0.1$  &$30$\\
        AzV\,175        &$4.97$ &$-7.55^{\dagger}$ &$437$ &$498$  &$1.0$  &$0.2$  &$30$\\
        Sk 191          &$5.67$ &$-6.81^{\dagger}$ &$381$  &$753$  &$2.5$  &$0.2$  &$30$\\
        AzV\,18         &$5.45$ &$-7.19$ &$388$  &$310$  &$3.5$  &$0.1$  &$25$\\
        NGC330-ELS-04   &$4.93$ &$-7.71^{\dagger}$ &$301$  &$490$  &$1.0$  &$0.2$  &$25$\\
        AzV\,187        &$5.37$ &$-7.77^{\dagger}$ &$307$  &$627$  &$1.0$  &$0.2$  &$25$\\
        AzV\,22         &$5.22$ &$-6.75$ &$160$  &$220$  &$2.5$  &$0.1$  &$25$\\
        AzV\,393        &$5.53$ &$-5.94$ &$237$  &$206$  &$2.5$  &$0.1$  &$25$\\
        AzV\,343        &$4.65$ &$-8.20^{\dagger}$ &$200$  &$244$  &$1.0$  &$0.1$  &$25$\\
        AzV\,324        &$4.70$  &$-8.05^{\dagger}$ &$178$  &$310$  &$1.0$  &$0.1$  &$25$\\
         \hline
        \end{tabular}
        \tablefoot{$\log{\dot{M}}$ is subject to an average uncertainty of 0.4~dex. $\varv_{\rm esc}$ is the escape velocity at the photosphere.}
    \end{table}
In Table~\ref{table:wind}, we present the derived wind parameters for our sample. The diverse nature of our sample is shown in the variety of $\log{\dot{M}/M_{\odot}\,{\rm yr}^{-1}}$, covering a range of $\approx-8.20$ to $-5.94$~dex. The highest mass-loss rate is attributed to the mid-B hypergiant AzV\,393, and the lowest to the late-B supergiant AzV\,343. 

\begin{figure*}
    \centering
     \includegraphics[width=\hsize]{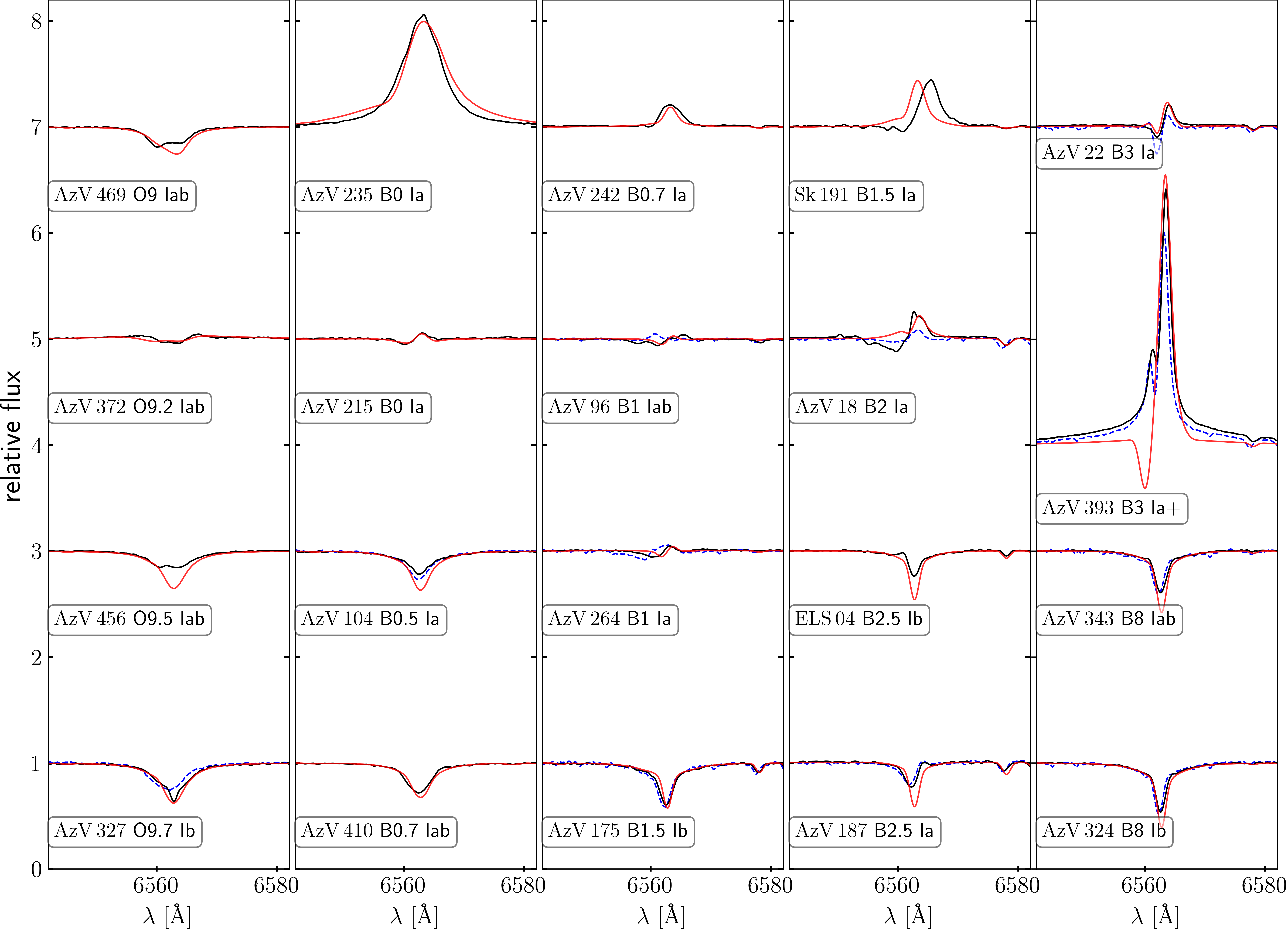}
         \caption{Best fits (solid red lines) to XShootU H$\alpha$ (+ $\ion{C}{II}~\lambda6578$) profiles (solid black lines). Magellan/MIKE spectra are also shown (dashed blue lines), where available, to illustrate line variability.}
         \label{Halpha_all}
    \end{figure*}

\begin{figure}
    \centering
     \includegraphics[width=\hsize]{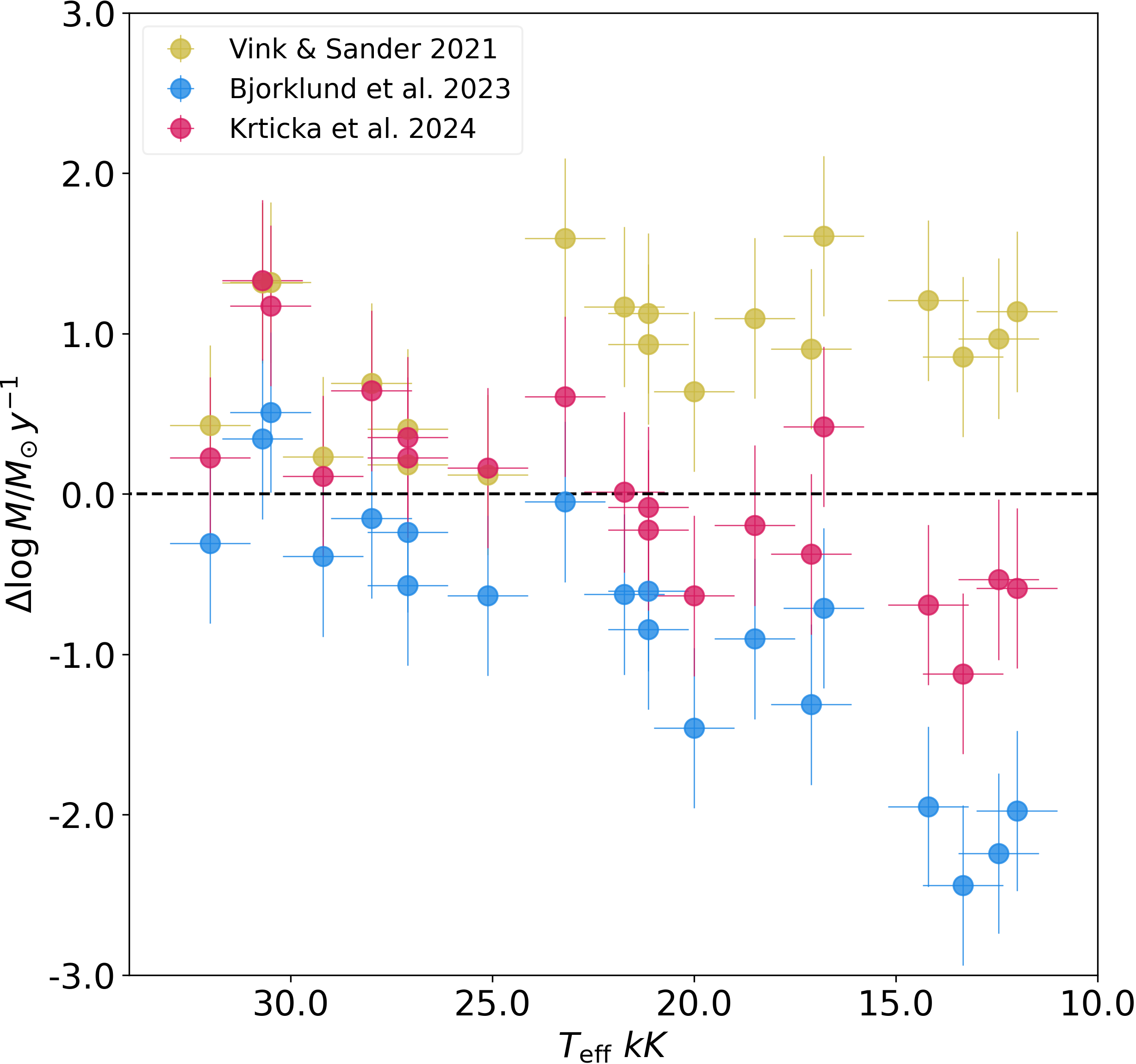}
         \caption{$\Delta\log{\dot{M}}$ vs $T_{\rm eff}$. $\Delta\log{\dot{M}}$ is the difference between our derived $\Delta\log{\dot{M}}$ and those obtained from numerical recipes from \citet{vinksander2021} (yellow dots), \citet{bjorklund2023} (blue dots), and \citet{krticka2024} (red dots).}
         \label{num_Mdot}
    \end{figure}
In Fig.~\ref{Halpha_all}, we present our best fits to (red solid line) H$\alpha$ profiles from XShootU spectra \citep[DR1]{Sana2024}. We also include H$\alpha$ profiles from MIKE observations \citep{mike} to illustrate the typical variability of this wind line. We quantify the $\log{\dot{M}}$ variability via test fits to MIKE H$\alpha$ observations, and obtain $\Delta\log{\dot{M}/M_{\odot}\,{\rm yr^{-1}}}=~\pm0.15$, which was taken into account when calculating the uncertainties of $\log{\dot{M}}$. 

We obtain satisfactory fits for cases in which the H$\alpha$ is fully or partially in emission, except for Sk\,191, the H$\alpha$ of which shows a redward radial velocity shift that is not present in other photospheric lines. In the cases in which H$\alpha$ is fully in absorption, we focus on fitting the wings. In any case, when determining $\dot{M}$, we take into account the level of saturation of P Cygni profiles. Nevertheless, for cases in which H$\alpha$ is fully in absorption, we consider the $\log{\dot{M}}$ presented in Table~\ref{table:wind} as an upper limit on mass loss.

In Fig.~\ref{num_Mdot}, we present a comparison between our derived $\log{\dot{M}}$ and the mass-loss rates predicted by the numerical recipe of \citet{vinksander2021}, \citet{bjorklund2023}, and \citet{krticka2024}, assuming $Z=0.2~Z_{\odot}$. We find that for $T_{\rm eff}\gtrapprox 22~{\rm kK}$ all recipes perform well and match our derived $\log{\dot{M}}$ within the uncertainties. For $T_{\rm eff}\lessapprox22~{\rm kK}$, which is believed to be the region where the predicted bi-stability jump occurs \citep{vink2001}, the $\log{\dot{M}}$ from the recipe of \citet{vinksander2021} predictably become much higher than ours by $\approx~1.5$~dex. The recipe of \citet{krticka2024} continues to perform well on the cool side of the bi-stability jump, whereas the values from the recipe of \citet{bjorklund2023} gradually become much smaller than our measurements. 

In general, $\dot{M}$ predictions from numerical recipes agree reasonably well with each other and with empirically derived $\dot{M}$ in the O star regime, but deviate greatly from one another and from empirical $\dot{M}$ in the B star regime. Recalling the discussion of the evolutionary status of our sample in Section~\ref{sec:HRD}, we conclude that predictions from numerical recipes are valid for stars on the MS, where massive stars spend most of their lives and where most mass loss takes place. However for post-MS stars, the empirical mass loss rates behave differently than what is predicted by these numerical recipes.

\begin{figure}
    \centering
     \includegraphics[width=\hsize]{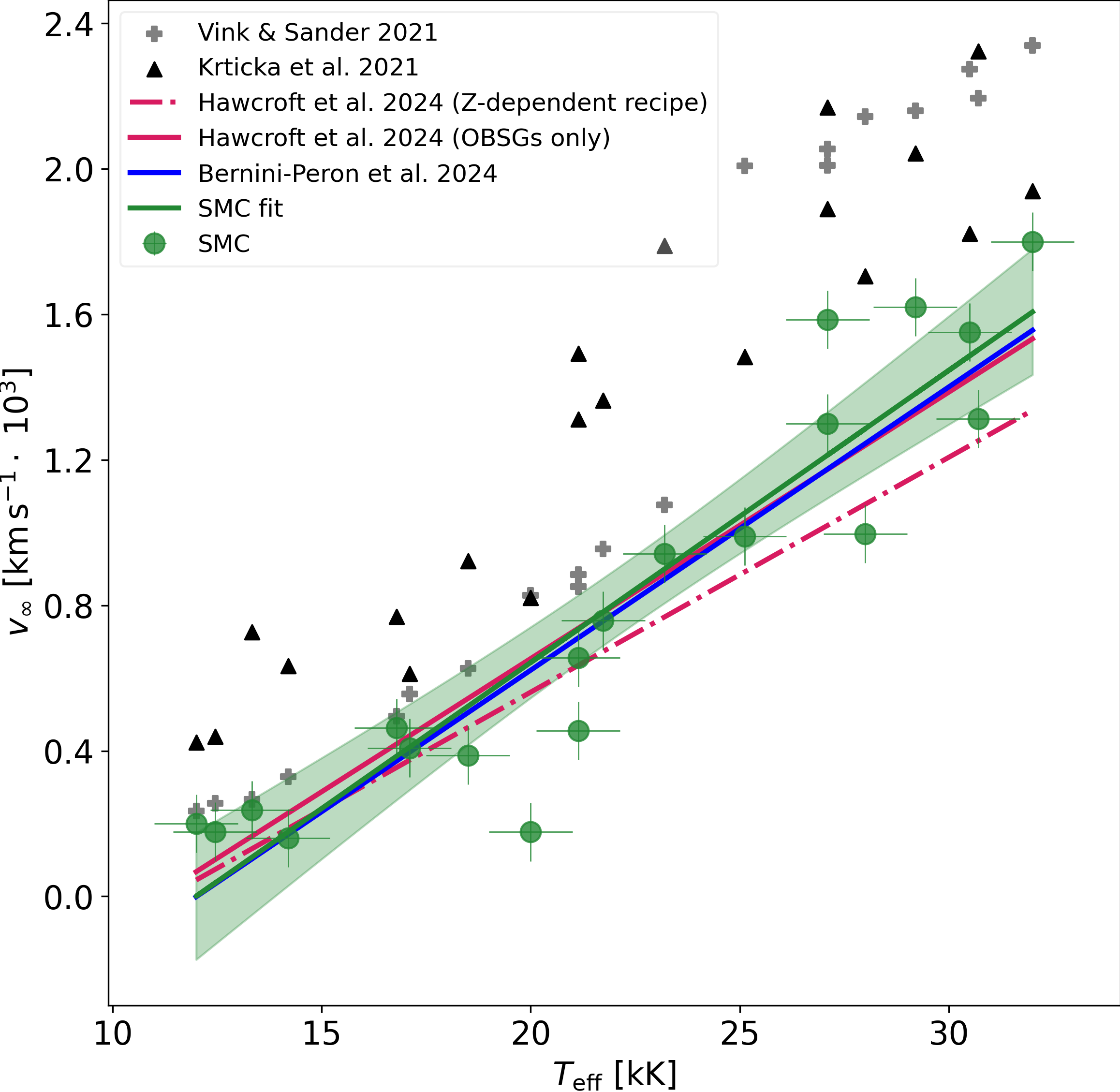}
         \caption{$\varv_{\infty}$ vs $T_{\rm eff}$. The green dots represent our results. The green line is the linear fit. The dashed magenta line is the Z-dependent $\varv_{\infty}$-$T_{\rm eff}$ relation from \citet{hawcroft2024}. The solid magenta line was obtained from fitting \citet{hawcroft2024} results for O and B supergiants only. The solid blue line was obtained from fitting the results of \citet{bernini2024}. The grey crosses are $\varv_{\infty}$ calculated from \citet{vinksander2021} recipe. The black triangles  represent $\varv_{\infty}$, calculated from the recipe in \citet{krticka2021}.}
         \label{vinf_comp}
    \end{figure}
    
Our sample spans a wide range of $\varv_{\infty} \approx 160$--$1800~{\rm km\,s^{-1}}$. In Fig.~\ref{vinf_comp} we show $\varv_{\infty}$ versus $T_{\rm eff}$, with our results compared to the empirical recipe of \citet{hawcroft2024}, results from \citet{bernini2024}, and velocities calculated from the numerical recipes of \citet{vinksander2021} and \citep{krticka2021}. We find a difference in the slopes with similar offsets. The main caveat of this comparison is that \citet{hawcroft2024} obtained their results by employing the SEI method, fitting only the $\ion{C}{IV}~\lambda1548$ P Cygni for stars no later than B1 and with various luminosity classes. $\varv_{\infty}$ prediction for $T_{\rm eff}$ values below $21~{\rm kK}$, therefore, should not be considered as reliable. 

We fitted our results with a simple linear fit of the form
\begin{equation}
  \label{eq:v-t} 
   \varv_{\infty} [{\rm km\,s^{-1}}] = aT_{\rm eff} [{\rm kK}]~-~ b[{\rm km\,s^{-1}}].
\end{equation}
\begin{table}
        \caption{Fitting parameter a (slope) and b (intercept) for Equation~\ref{eq:v-t}.}
        \label{table:v-t}    
        \def\arraystretch{1.1}  
        \centering      
        \small                               
        \addtolength{\tabcolsep}{-0.1em}
        \begin{tabular}{c c c}        
        \hline\smallskip
        study                                  &a              &b\\
                                               &kK             &${\rm km\,s^{-1}}$\\
        \hline
        This study                             &$0.08\pm0.01$  &$963\pm200$\\
        \citep[][entire sample]{hawcroft2024}  &$0.09\pm0.01$  &$1560\pm420$\\
        \citep[][OB SGs only]{hawcroft2024}    &$0.07\pm0.01$  &$811\pm300$\\
        \citep{bernini2024}                    &$0.08\pm0.02$  &$936\pm350$\\
        \hline
        \end{tabular}
    \end{table}
In Table~\ref{table:v-t}, we present the slope, $a$, and intercept, $b$, from Equation~\ref{eq:v-t}, and compare them to the results of \citet{hawcroft2024} and \citet{bernini2024}. We find that our derived fitting parameters are in excellent agreement with the parameters from \citep{hawcroft2024} when considering only O and B supergiants. We also find excellent agreement with the results of \citet{bernini2024}. 

Our velocities agree with predictions from the recipe of \citet{vinksander2021} below $25~{\rm kK}$, but are lower by $\approx 30\%$ for temperatures above that threshold. We also find that the recipe from \citet{krticka2021} systematically overpredicts the velocities by an average of $\approx500~{\rm km\,s^{-1}}$.

Our results in Table~\ref{table:wind} show that a high $\beta$ ($\approx 2.5\pm0.6$) is preferred to obtain a satisfactory fit for H$\alpha$ in the cases in which it is fully or partially in emission. This agrees with the findings of \citet{bernini2024}, who require $\beta>2$ for most of their SMC sample. Similarly, \citet{crowther2006} obtained an average $\beta=2$ from a sample of early Galactic B supergiants.

Due to the weakness of the winds of SMC supergiants, and consequently the lack of broad saturated P Cygni profiles in the UV and H$\alpha$ emission, we were able to constrain $f_{\rm vol,\infty}$ for only nine stars in our sample. We did not find any correlation between $f_{\rm vol,\infty}$ and $T_{\rm eff}$, $L_{\rm bol}$, or $\dot{M}$. This is similar to our findings in Paper~XIII, and has been the case in recent studies conducting UV and optical spectroscopic analysis of OB stars in multiple environments and utilising various codes with different clumping implementations \citep{hawcroft2024a, bernini2024, verhamme2024, brands2025}. 

\subsection{Rotation} 
In Table~\ref{table:2} we present $\varv_{\rm rot}\sin{i}$ of our sample. Our sample spans a range of $\varv_{\rm rot}\sin{i}$ from $25~{\rm km\,s^{-1}}$ to $100~{\rm km\,s^{-1}}$, with a mean $\varv_{\rm rot}\sin{i}$ of $51~{\rm km\,s^{-1}}$ and a standard deviation of $19~{\rm km\,s^{-1}}$. This agrees with the findings of \citet{dufton2006}, who analysed a sample of 24 SMC and Galactic B supergiants and found a linear correlation between $\varv_{\rm rot}$ and $T_{\rm eff}$ with a value of $\varv_{\rm rot}\sin{i}$ of $\approx60$ and $30~{\rm km\,s^{-1}}$ for $T_{\rm eff}=28$ and $12~{\rm kK}$, respectively.

\begin{table}
  \caption{Comparison of our adopted broadening parameters with those obtained via \textsc{IACOB-BROAD} \citep{simon-diaz2014} applied to high-resolution MIKE data.} 
  \def\arraystretch{1.5}
  \label{table:rot}    
  \centering      
  \small                               
  \addtolength{\tabcolsep}{-0.25em}
  \begin{tabular}{c c c | c c c }          
      \hline\hline{\smallskip}
Star                         &\multicolumn{2}{c}{by eye+XShootU}             &\multicolumn{3}{c}{\textsc{IACOB-BROAD+MIKE}}\\
      \hline{\smallskip}
                &$\varv_{\rm rot}\sin{i}$     &$\varv_{\rm mac}$   &$\varv_{\rm rot}\sin{i}$  &$\varv_{\rm mac}$     &line\\
                &${\rm km\,s^{-1}}$           &${\rm km\,s^{-1}}$  &${\rm km\,s^{-1}}$        &${\rm km\,s^{-1}}$    & \\ 
\hline
AzV\,327        &$55$                         &$20$                &$64^{+35}_{-62}$          &$87^{+42}_{-55}$      &$\ion{Si}{III}~\lambda4552$\\
AzV\,104        &$65$                         &$20$                &$80^{+19}_{-53}$          &$45^{+63}_{-43}$      &$\ion{Si}{III}~\lambda4552$\\
AzV\,96         &$50$                         &$20$                &$52^{+16}_{-24}$          &$59^{+23}_{-29}$      &$\ion{Si}{III}~\lambda4552$\\
AzV\,264        &$45$                         &$20$                &$45^{+15}_{-43}$          &$33^{+39}_{-31}$      &$\ion{Si}{III}~\lambda4552$\\
AzV\,18         &$50$                         &$20$                &$43^{+11}_{-18}$          &$44^{+20}_{-21}$      &$\ion{Si}{III}~\lambda4552$\\
NGC330-ELS-4    &$35$                         &$20$                &$33^{+9}_{-18}$           &$29^{+19}_{-19}$      &$\ion{Si}{III}~\lambda4552$\\
AzV\,187        &$40$                         &$20$                &$36^{+17}_{-34}$          &$41^{+27}_{-35}$      &$\ion{Si}{III}~\lambda4552$\\
AzV\,393        &$30$                         &$20$                &$35^{+16}_{-32}$          &$30^{+33}_{-28}$      &$\ion{Si}{III}~\lambda4552$\\
AzV\,343        &$35$                         &$20$                &$44^{+9}_{-21}$           &$19^{+33}_{-17}$      &$\ion{Si}{II} ~\lambda6347$\\
AzV\,324        &$25$                         &$20$                &$21^{+6}_{-9}$            &$26^{+8}_{-10}$       &$\ion{Si}{II} ~\lambda6347$\\
    \noalign{\smallskip}
      \hline
  \end{tabular}
\end{table}

In Table~\ref{table:rot}, we compare our derived $\varv_{\rm rot}\sin{i}$ to the $\varv_{\rm rot}\sin{i}$ and $\varv_{\rm mac}$ that were obtained from the \textsc{IACOB-BROAD} tool, which employs a combined Fourier transform and goodness-of-fit approach that allows for the extraction of line-broadening parameters \citep{simon-diaz2014}. We applied \textsc{IACOB-BROAD} to a subsample of our stars with high-resolution MIKE data. The line we used to extract the rotational properties is $\ion{Si}{III}~\lambda4552$ for the entire subsample, except for the late B supergiants AzV\,324 and AzV\,343, for which we used $\ion{Si}{II}~\lambda6347$. 

We find that our derived $\varv_{\rm rot}\sin{i}$ agree with those obtained from \textsc{IACOB-BROAD} within $10~{\rm km\,s^{-1}}$, except for AzV\,104 and AzV\,327, where the \textsc{IACOB-BROAD} $\varv_{\rm rot}\sin{i}$ is $15$ and $20~{\rm km\,s^{-1}}$ higher than ours, respectively. On the other hand, we tend to underestimate $\varv_{\rm mac}$ on average by $\approx20~{\rm km\,s^{-1}}$. The largest difference in $\varv_{\rm mac}$ are observed in AzV\,327 and AzV\,96, for which the values obtained from \textsc{IACOB-BROAD} are higher by $67$ and $39~{\rm km\,s^{-1}}$, respectively.

\subsection{Chemical composition}
\begin{table}
        \caption{Best-fitting helium (Y, by mass) and CNO ($\epsilon_{\rm X}$, by number) abundances.}
        \label{table:abund}    
        \def\arraystretch{1.1}  
        \centering      
        \small                               
        \addtolength{\tabcolsep}{-0.1em}
        \begin{tabular}{c c c c c c c}        
        \hline\smallskip
        Star    &Y &$\epsilon_{\rm C}$&$\epsilon_{\rm N}$&$\epsilon_{\rm O}$&$\frac{\Sigma {\rm CNO}}{\Sigma {\rm CNO}_{\rm SMC}}$  &$\frac{\Sigma {\rm CNO}}{\Sigma {\rm CNO}_{\odot}}$\\
        \hline
        AzV\,469        &$0.52$ &$7.3$ &$8.1$ &$8.2$ &$2.2$ &$0.42$\\
        AzV\,372        &$0.52$ &$7.2$ &$7.8$ &$8.2$ &$1.7$ &$0.33$\\
        AzV\,456        &$0.30$ &$7.2$ &$7.2$ &$8.1$ &$1.0$ &$0.19$\\
        AzV\,327        &$0.46$ &$7.2$ &$7.8$ &$8.1$ &$1.3$ &$0.24$\\
        AzV\,235        &$0.30$ &$7.0$ &$7.8$ &$7.9$ &$1.0$ &$0.19$\\
        AzV\,215        &$0.30$ &$7.0$ &$8.0$ &$8.1$ &$1.6$ &$0.31$\\
        AzV\,104        &$0.30$ &$7.0$ &$7.7$ &$8.0$ &$1.0$ &$0.20$\\
        AzV\,410        &$0.37$ &$7.1$ &$7.7$ &$8.0$ &$1.1$ &$0.22$\\
        AzV\,242        &$0.30$ &$7.0$ &$7.5$ &$8.0$ &$1.0$ &$0.18$\\
        AzV\,96         &$0.30$ &$6.7$ &$7.9$ &$7.8$ &$0.9$ &$0.18$\\
        AzV\,264        &$0.46$ &$6.8$ &$7.9$ &$7.8$ &$1.1$ &$0.21$\\
        AzV\,175        &$0.30$ &$7.1$ &$7.5$ &$8.1$ &$1.1$ &$0.21$\\
        Sk 191          &$0.43$ &$6.9$ &$7.9$ &$8.0$ &$1.2$ &$0.23$\\
        AzV\,18         &$0.43$ &$7.1$ &$7.7$ &$8.0$ &$1.1$ &$0.22$\\
        NGC330-ELS-04   &$0.40$ &$7.0$ &$7.9$ &$7.9$ &$1.2$ &$0.23$\\
        AzV\,187        &$0.30$ &$7.2$ &$7.5$ &$8.0$ &$0.9$ &$0.18$\\
        AzV\,22         &$0.46$ &$6.9$ &$8.0$ &$8.0$ &$1.4$ &$0.27$\\
        AzV\,393        &$0.30$ &$6.9$ &$7.8$ &$8.0$ &$1.1$ &$0.21$\\
        AzV\,343        &$0.40$ &$7.6$ &$7.9$ &$7.5^{*}$ &$1.1$ &$0.20$\\
        AzV\,324        &$0.34$ &$7.2$ &$7.6$ &$7.5^{*}$ &$0.6$ &$0.11$\\
         \hline
        Averages        &$0.37$ &$7.0$ &$7.8$ &$7.9$ &$1.1$ &$0.23$\\
        \hline
        \end{tabular}
        \tablefoot{The uncertainties of the helium mass fraction are $10\%$ . The adopted uncertainty of CNO abundances is $\pm0.3$~dex. The last two columns represent the cumulative CNO mass fraction relative to the cumulative CNO SMC baseline and solar baseline, respectively. Oxygen abundance values with the asterisk subscript were obtained from fitting the $\ion{O}{I}~\lambda\lambda\lambda7772-7774-7775$ multiple, and are considered lower limits.}
    \end{table}

\begin{figure}
    \centering
     \includegraphics[width=\hsize]{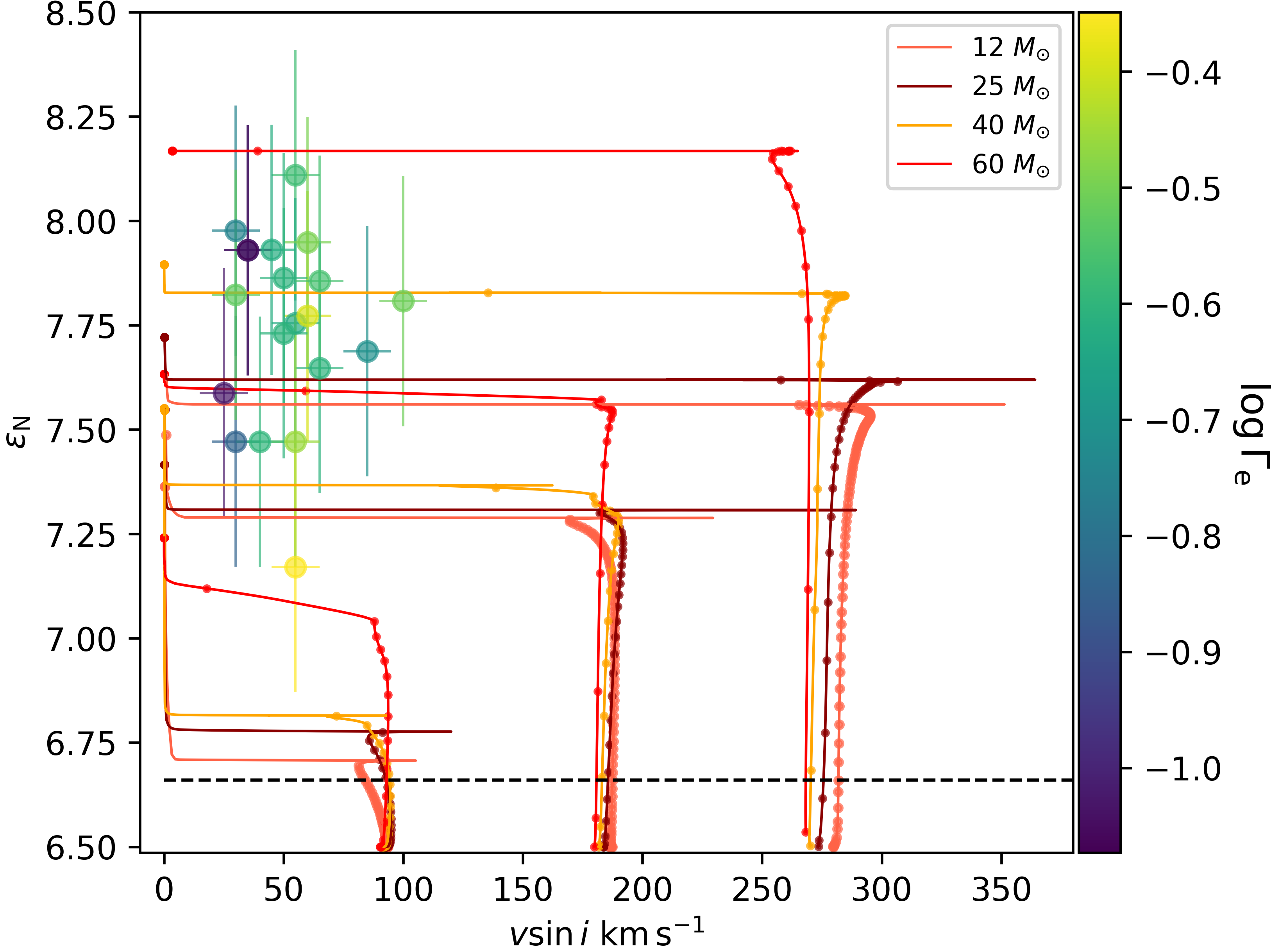}
         \caption{$\epsilon_{\rm N}$ vs $\varv_{\rm rot}\sin{i}$. The colour scheme corresponds to the value of $\Gamma_{\rm e}$. The solid pink, brown, orange, and red lines represent SMC evolutionary tracks for initial masses of $12~M_{\odot}$, $25~M_{\odot}$, $40~M_{\odot}$, and $60~M_{\odot}$, respectively, with initial rotational velocities of $~110,~250,~350~{\rm km\,s^{-1}}$ \citep{brott2011}. We multiplied by a factor of $\pi/4$ to take into account the inclination of the rotation axis \citep{hunter2008}. The dashed black line represents the baseline $\epsilon_{\rm N}$ in the SMC \citep{xshootu1}. The dots along the evolutionary tracks represent time steps of $0.1~{\rm Myr}$}
         \label{N_vsini}
    \end{figure}
    
We present the best-fitting chemical abundances for our sample in Table~\ref{table:abund}. We find moderate helium enhancement for our sample relative to the SMC baseline \citep[Y$\sim$0.25,][]{Russell&Dopita1990}, which is to be expected in a sample comprised of blue supergiants. Fig.~\ref{N_vsini} shows that all stars in our sample exhibit significant nitrogen enrichment. According to the rotating single-star evolutionary models of \citet{brott2011} that include rotational mixing, this enrichment can be explained by the high initial rotational velocities ($\varv_{\rm rot}\sin{i}>300~{\rm km\,s^{-1}}$), which cause nitrogen synthesised in the interior of the star to be transported to the surface. However, \citet{dufton2006} found that $\varv_{\rm rot}\sin{i}$ of SMC B supergiants falls in the range of $\approx60$--$30~{\rm km\,s^{-1}}$, whereas \citet{mokiem2006} found that $\varv_{\rm rot}\sin{i}$ falls in the range of $\approx150$--$180~{\rm km\,s^{-1}}$ from a diverse sample of SMC O and early B stars. 

Spectroscopic studies of the rotational properties of O stars in the SMC, LMC, and the MW show that most early O stars, which are the progenitors of B supergiants, are not fast rotators. \citet{ramachandran2019} find that $\approx75\%$ of SMC O stars have $\varv_{\rm rot}\sin{i} < 200~{\rm km\,s^{-1}}$. In the Tarantula nebula, \citet{ramirez2013} find that the distribution of $\varv_{\rm rot}\sin{i}$ of O stars peaks at $\approx80~{\rm km\,s^{-1}}$. Similarly, \citep{holgado2022} find that the distribution of $\varv_{\rm rot}\sin{i}$ of O stars across all luminosity classes in the MW peaks at $\approx80~{\rm km\,s^{-1}}$. 

These observed distributions challenge the assumption that extremely high initial or current rotational velocities are the explanation for the nitrogen enhancement. Only about $\approx25\%$ of the massive stars are effectively single, with the overwhelming majority of the fastest rotators being the product of binary interaction \citep{demink2013, britavskiy2025, villasenor2025, bodensteiner2025, sana2025}. Therefore, other processes should be considered to explain the nitrogen enhancement, such as binary interaction in the history of the star or the potential existence of a currently undetected companion. 

\begin{figure}
    \centering
     \includegraphics[width=\hsize]{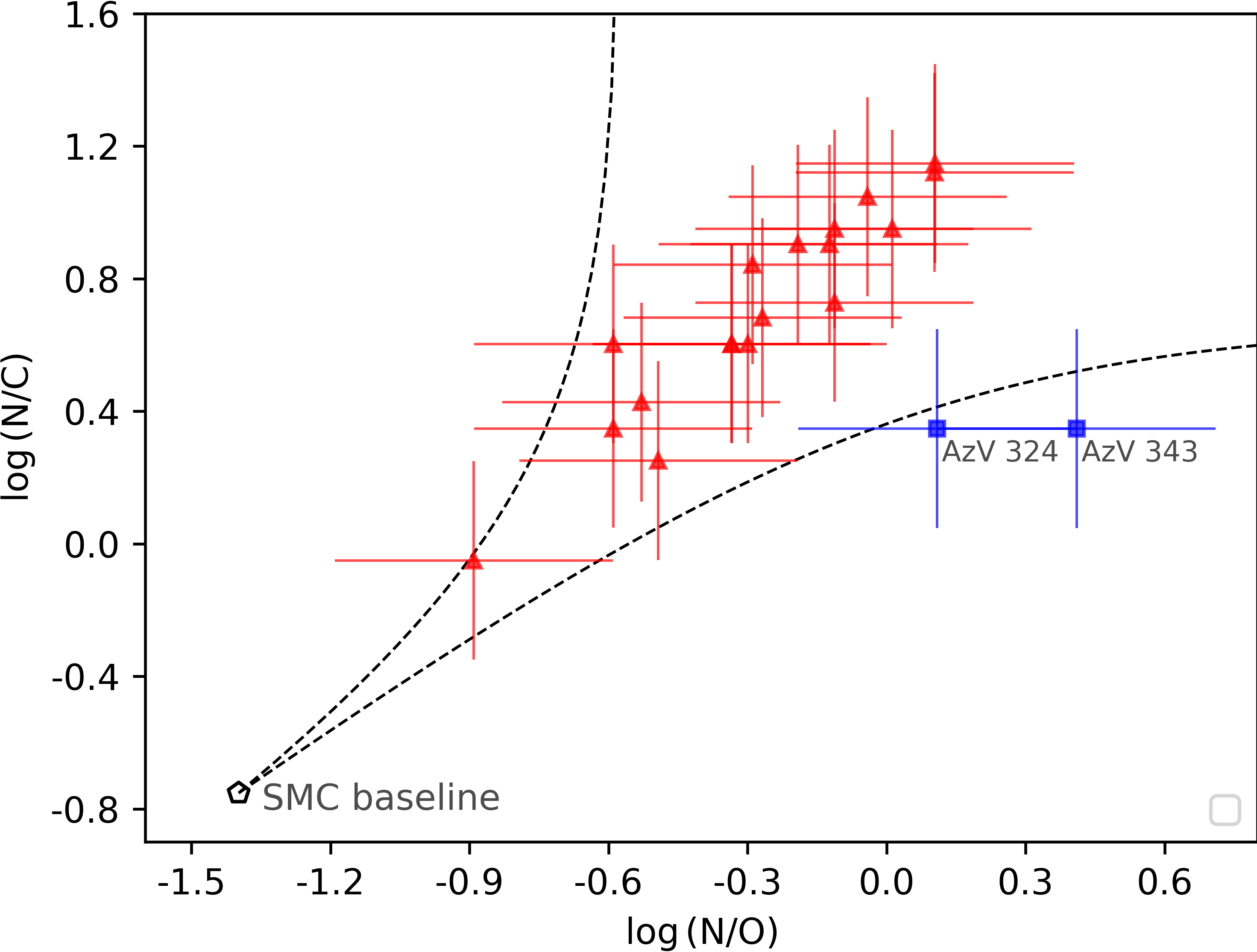}
         \caption{$\log{{\rm N}/{\rm C}}$ vs $\log{{\rm N}/{\rm O}}$. The red triangles show the distribution of our sample. The stars for which the oxygen abundance was obtained from fitting the $\ion{O}{I}~\lambda\lambda\lambda7772-7774-7775$ multiplet are shown in blue squares. The dashed black lines represent the upper and lower boundaries adopted from \citet{maeder2014}.}
         \label{CNO}
    \end{figure}
Fig.~\ref{CNO} shows the ratio N/C versus N/O. Following the analytical solution of \citet{maeder2014}, the upper and lower CNO limits (dashed black lines) were calculated using the baseline CNO abundance averages of the SMC from \citet{xshootu1}. The majority of our sample falls between the two boundaries and agrees with the expected yield of elements processed by the CNO cycle. The exceptions are AzV\,324 and AzV\,343, for which we obtained the oxygen abundance by fitting the $\ion{O}{I}~\lambda\lambda\lambda7772-7774-7775$ multiplet, due to the absence of any other oxygen lines in the optical range. We found that a satisfactory fit to this complex line requires a reduction in the oxygen mass fraction in the model by a factor of four, which explains the high N/O ratios in Fig.~\ref{CNO} for these objects. As is explained in Paper~XIII, the $\ion{O}{I}~\lambda\lambda\lambda7772-7774-7775$ multiplet is extremely sensitive to $T_{\rm eff}$ variations at this range; therefore, the oxygen content obtained by fitting those lines is highly uncertain.

Our sample has two stars (AzV\,469 and AzV\,327) in common with \citet{martins2024} , who employed \textsc{CMFGEN} to obtain the chemical abundances of a sample of SMC and LMC O stars using spectroscopy from ULLYSES and XShootU. We find that our obtained $\epsilon_{\rm N}$ for these stars agrees within $0.1$~dex with their derived nitrogen abundances, whereas carbon and oxygen do not agree so well, though they match within the quoted uncertainties.

We find that the total CNO mass fraction is conserved for most of our sample, except AzV\,469, AzV\,372, and AzV\,215. In the case of these stars, enhanced nitrogen abundances were required to obtain a satisfactory match to the diagnostic lines, and oxygen mass fractions slightly larger than baseline were also required to obtain satisfactory fits for oxygen lines. This explains the implausibly large ratio of cumulative CNO mass fraction to the cumulative solar CNO mass fraction, $\Sigma {\rm CNO}/\Sigma {\rm CNO}_{\odot}$. We use these example as a reminder of the large uncertainties associated with the determination of abundances in supergiants from our approach, which are in part due to the sensitivity of the ionisation structure of CNO lines to variations in $T_{\rm eff}$ and $\log{g}$, but also the fixed micro-turbulent velocity, and the extent of the atomic model adopted in the calculation (number of important levels and superlevels), and the incompleteness of the atomic data. 

\section{Discussion}
\label{sec:discussion}
\subsection{Metallicity effect}
\begin{figure}
    \centering
     \includegraphics[width=\hsize]{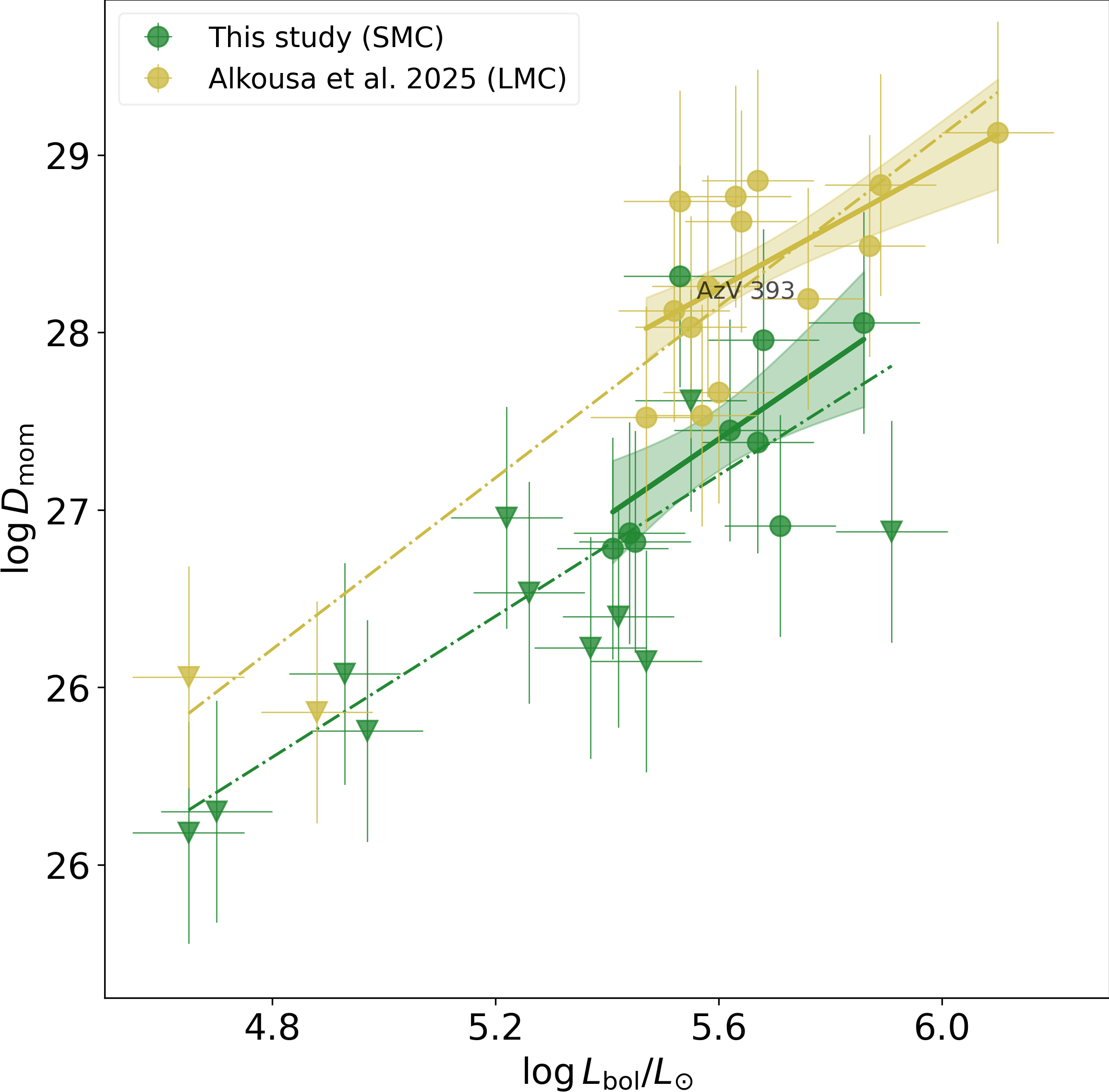}
         \caption{$D_{\rm mom}$ vs $log{L_{\rm bol}}$. The green symbols represent the SMC sample. The yellow symbols represent the LMC sample from Paper~XIII. Circles indicate stars with reliable UV and optical wind diagnostics. Triangles indicate upper limits. Solid lines are fits to the circles. Dash-dotted lines are fits for the entire sample.}
         \label{Dmom}
    \end{figure}

Fig.~\ref{Dmom} shows the modified wind momentum, which was introduced by \citet{kudritzki1999} as $D_{\rm mom}= \dot{M}\varv_{\infty}\sqrt{R_{\ast}\big/R_{\odot}}$, as a function of $L_{\rm bol}$. We fitted our derived values using the relation
\begin{equation}
    \label{eq:Dmom} 
      \log{D_{\rm mom}^{\rm SMC}} =  \log{D_{0}} + x\log{\frac{L_{\rm bol}}{L_{\odot}}}. 
\end{equation}
In Fig.~\ref{Dmom}, solid lines are fits to the stars (circles) that show both P Cygni profiles in the UV and H$\alpha$ in emission, making their determined wind parameters the most reliable. We can see that including stars with H$\alpha$ in absorption (triangles) in the linear fit slightly changes the slope of the fit for the LMC sample from Paper~XII (dash-dotted green line). Doing the same for our SMC sample decreases the vertical offset of the fit (dash-dotted yellow line), albeit very slightly.

\begin{table}
  \caption{Slopes, $x$, and vertical offsets, $\log{D_{0}}$, of Equation~\ref{eq:Dmom} of this study and from Paper~XIII.}
  \def\arraystretch{1.5}
  \label{table:dmom}      
  \centering      
  \small                                
  \addtolength{\tabcolsep}{-0.3em}
  \begin{tabular}{c c c c c c}        
      \hline\hline{\smallskip}
         Galaxy          &$x$         &$\log{D_{0}}$    &Sp. T. &Lum. Cl. &num. stars\\
\hline 
SMC            &$1.73\pm 1.00$       &$17.8\pm 5.62$   &O9-B3 &I &9\\  
LMC            &$1.39\pm 0.54$       &$20.42\pm 3.04$   &O9-B3 &I &14\\ 
\hline 
    \noalign{\smallskip}
      \hline
  \end{tabular}
\end{table}

In Table~\ref{table:dmom}, we present the best fitting parameters of Equation~\ref{eq:Dmom} (slopes, $x$, and offsets, $\log{D_{0}}$). Comparing the $D_{\rm mom}$ of the current SMC sample to the LMC sample analysed in Paper~XIII, we find that the slope of the SMC is slightly steeper than the LMC, albeit quite similar. We also find a large difference in the intercepts between the two samples, which can be attributed to the $Z$ dependence of wind properties. 

\begin{figure}
    \centering
     \includegraphics[width=\hsize]{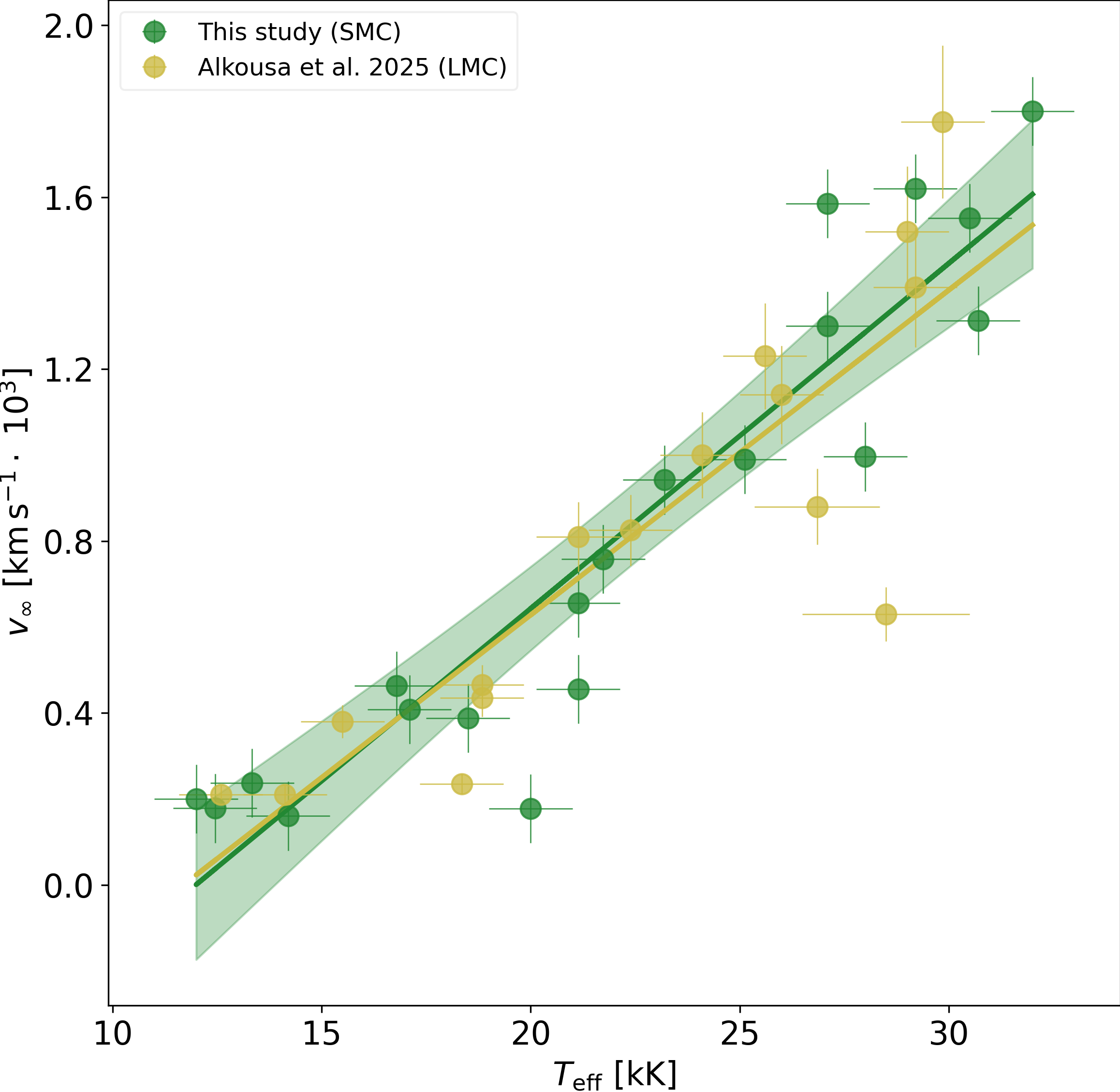}
         \caption{$\varv_{\infty}$ vs $T_{\rm eff}$. The symbol and colour encoding are the same as in Fig.~\ref{Dmom}.}
         \label{vinf_teff_Z}
    \end{figure}
\begin{figure}
    \centering
     \includegraphics[width=\hsize]{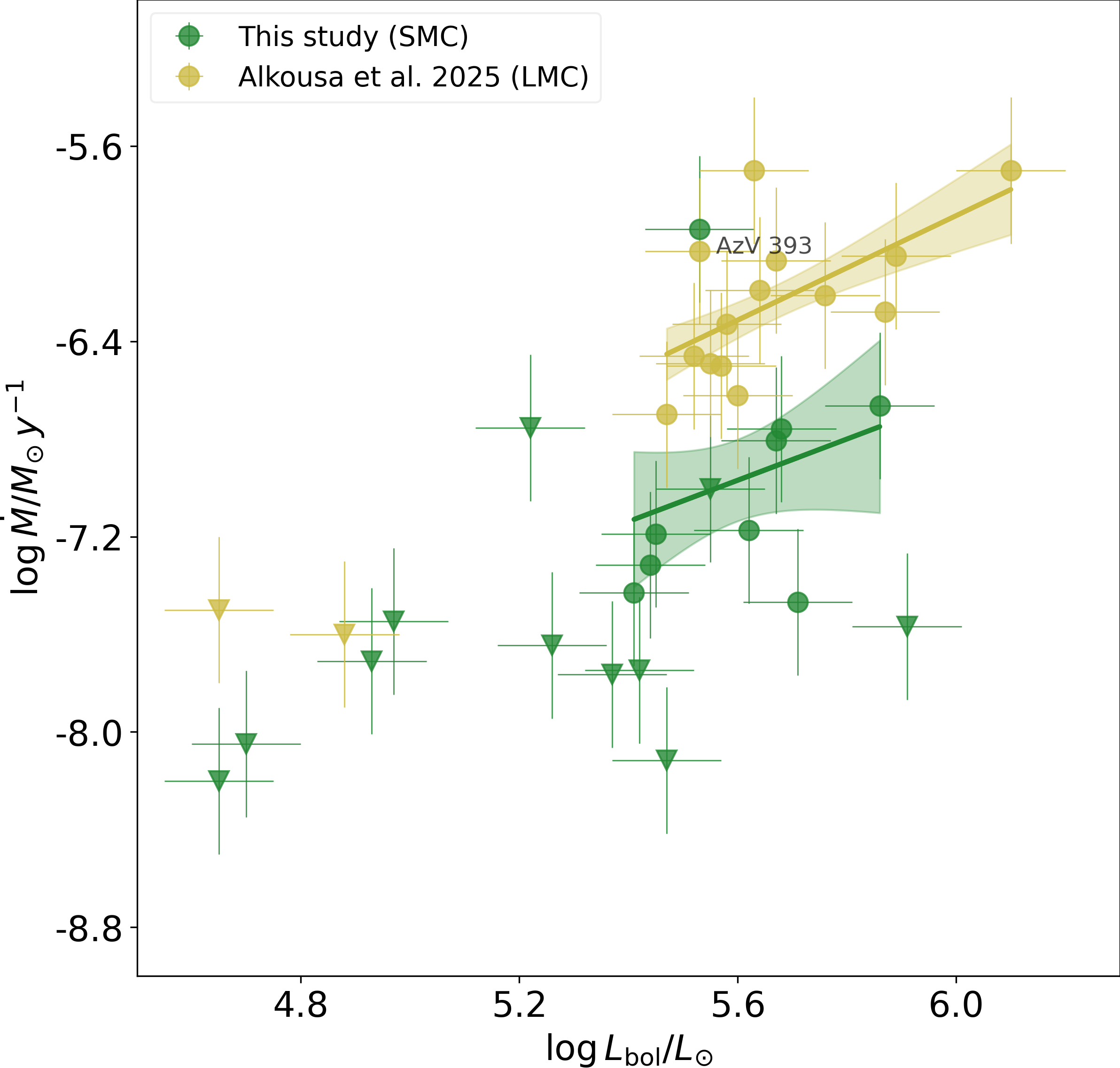}
         \caption{$\log{\dot{M}}$ vs $\log{L_{\rm bol}}$. The symbol and colour encoding are the same as in Fig.~\ref{Dmom}.}
         \label{Mdot_Lbol_Z}
    \end{figure}
In Figure~\ref{vinf_teff_Z}, we show $\varv_{\infty}$ as a function of $T_{\rm eff}$ of the present study compared to the results from Paper~XIII. Figure~\ref{Mdot_Lbol_Z} shows $\log{\dot{M}}$ vs $\log{L_{\rm bol}}$ from this study compared to Paper~XIII. $\varv_{\infty}$ does not show signs of Z dependence, whereas $\log{\dot{M}}$ shows a clear offset between the two samples, indicating a strong Z dependence. Therefore, we can assume that the Z dependence of $D_{\rm mom}$ is dominated by the Z dependence of $\dot{M}$.

In parallel with our study, \citet{verhamme2025} have recently undertaken an analysis of ULLYSES/XShootU SMC B supergiants. To investigate the effect of metallicity, they compare their analysis to an earlier study of LMC B supergiants \citep{verhamme2024}. Despite our overlapping samples, they arrived at different conclusions, which is discussed further in Appendix~\ref{Verhamme}.

Following \citet{krticka&kubat2018} and \citet{backs2024}, we obtained an equation for the $Z$-dependent $\log{D_{\rm mom}}$ of the form 
\begin{equation}
    \label{eq:Dmom_Z} 
      \log{D_{\rm mom}} = (a+b\log{\frac{Z}{Z_{\odot}}}) \log{\frac{L_{\rm bol}}{10^{6}L_{\odot}}} + c \log{\frac{Z}{Z_{\odot}}} + d,
\end{equation}
\begin{table}
  \caption{Fitting parameter of Equation~\ref{eq:Dmom_Z}.}  
  \def\arraystretch{1.5}
  \label{table:dmom_z}      
  \centering      
  \small                                
  \addtolength{\tabcolsep}{0.0em}
  \begin{tabular}{c c c}        
    \hline\hline{\smallskip}
    parameter   &This study &Backs et al. (2024) \\
    \hline
    a           &$1.16  \pm0.48$       &$0.64$    \\            
    b           &$-0.75 \pm1.14$      &$-2.84$   \\           
    c           &$1.38  \pm4.03$       &$0.71$    \\            
    d           &$29.17 \pm6.39$      &$29.27$   \\           
    \noalign{\smallskip}
      \hline
  \end{tabular}
\end{table}
where from Equation~\ref{eq:Dmom} $x = a+b\log{\frac{Z}{Z_{\odot}}}$, $\log{D_{0}} = c \log{\frac{Z}{Z_{\odot}}} + d$, and $a$, $b$, $c$, and $d$ are fitting parameters. The results of the fitting are presented in Table~\ref{table:dmom_z}, assuming that $Z_{\rm LMC} = 0.5 Z_{\odot}$ and $Z_{\rm SMC} = 0.2 Z_{\odot}$. We also include values from \citet{backs2024} in Table~\ref{table:dmom_z}.

\subsection{Bi-stability jump}
\begin{figure}[!htbp]
    \centering
     \includegraphics[width=\hsize]{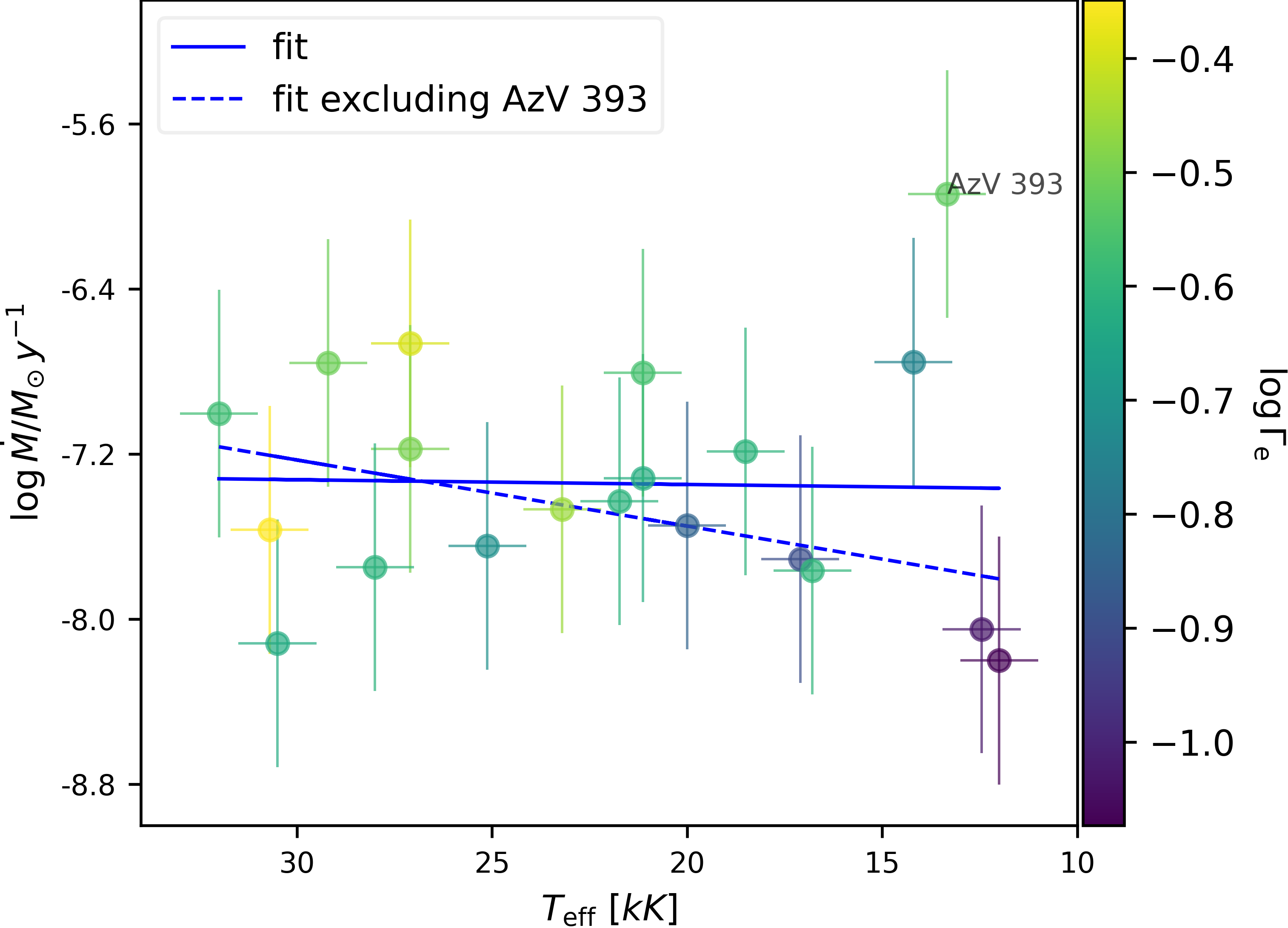}
         \caption{$\log{\dot{M}}$ versus $T_{\rm eff}$. The solid blue line is a simple linear fit to the entire sample. The colour gradient correlates with the
         value of $\Gamma_e$.}
         \label{Mdot_teff}
    \end{figure}
    
In Fig.~\ref{Mdot_teff}, we present our derived $\log{\dot{M}}$ in terms of $T_{\rm eff}$. This confirms that AzV\,393 possesses much higher $\dot{M}$, not just compared to objects with similar $T_{\rm eff}$, but even compared to the hotter OB supergiants in our sample. We do not find any sign of a large increase in $\dot{M}$ around 25-21~kK, which could be attributed to the bi-stability jump. Using a simple linear regression, we find that $\log{\dot{M}}$ is constant with $T_{\rm eff}$ (solid blue line). Excluding the hypergiant AzV\,393 from the fit (dashed blue line), we find that $\log{\dot{M}}$ decreases with $T_{\rm eff}$. 

\begin{figure}
    \centering
     \includegraphics[width=\hsize]{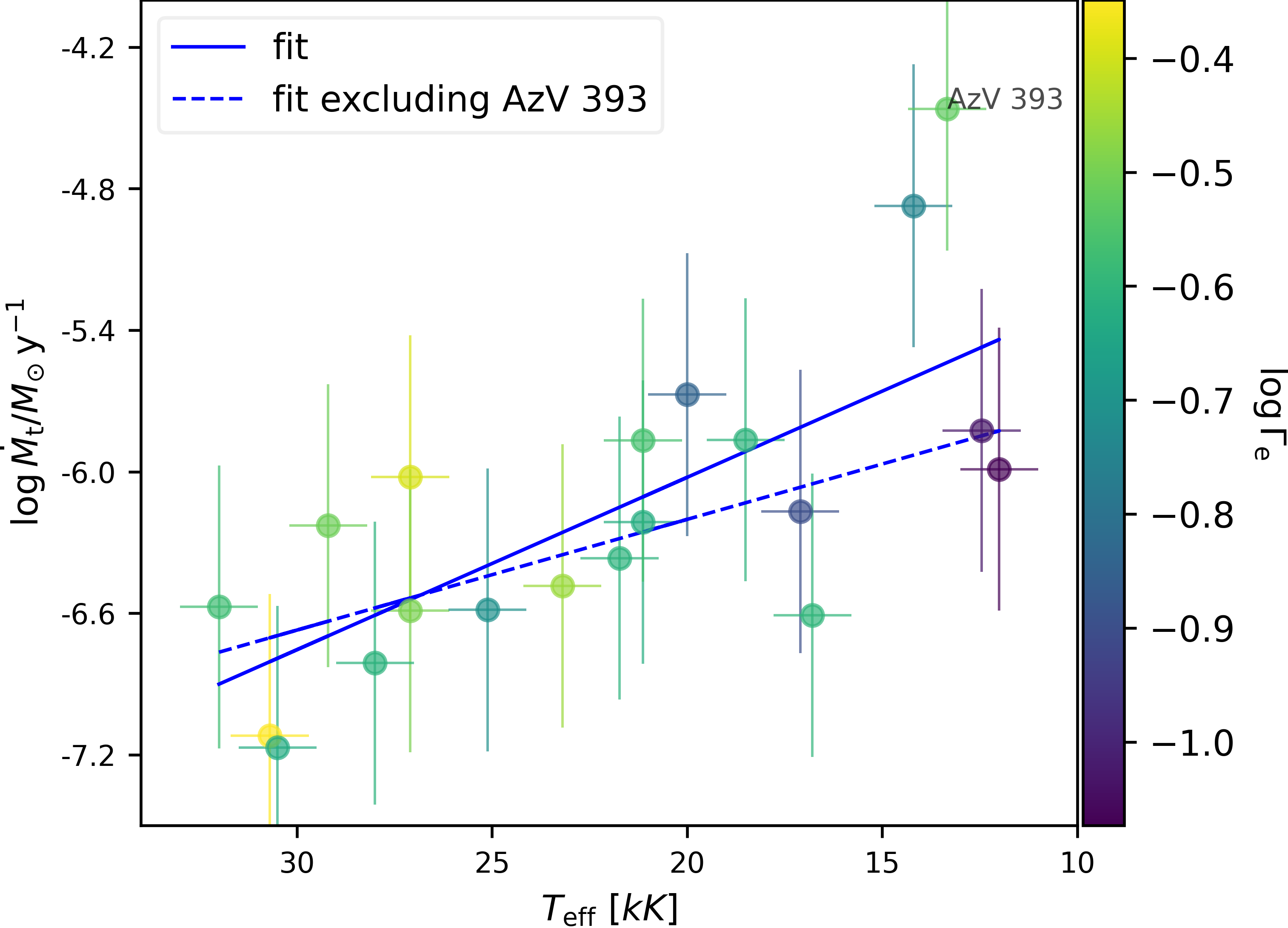}
         \caption{$\log{\dot{M}_{\rm t}}$ versus $T_{\rm eff}$. The symbol and colour encoding are the same as in Fig.~\ref{Mdot_teff}.}
         \label{transMdot_teff}
    \end{figure}

To investigate the effect of luminosity, we calculated the transformed mass-loss rate, which is defined as 
\begin{equation}
\label{eq:transMdot}
    \dot{M}_{\rm t} = \dot{M}\cdot f_{\rm vol,\infty}^{-1/2}\cdot(10^{3}{\rm km\,s^{-1}}/ \varv_{\infty})\cdot (10^{6} L_{\odot}/L_{\rm bol})^{3/4}.
\end{equation}
In Fig.~\ref{transMdot_teff}, we compare $\log{\dot{M}_{\rm t}}$ to $T_{\rm eff}$. We find that when including or excluding AzV\,393 in the linear fit, $\log{\dot{M}_{\rm t}}$ increases with temperature. This is in contrast to the results of \citet{bernini2024}, which hint at a constant behaviour of the mass-loss rate with temperature. The reason for this trend could be our comparably low values of $f_{\rm vol,\infty} < 0.2$ (i.e. a high degree of clumping in the winds), which lowers the mass-loss rates required to obtain a satisfactory fit to H$\alpha$ and to the UV P Cygni profiles. 

\begin{figure}
    \centering
     \includegraphics[width=\hsize]{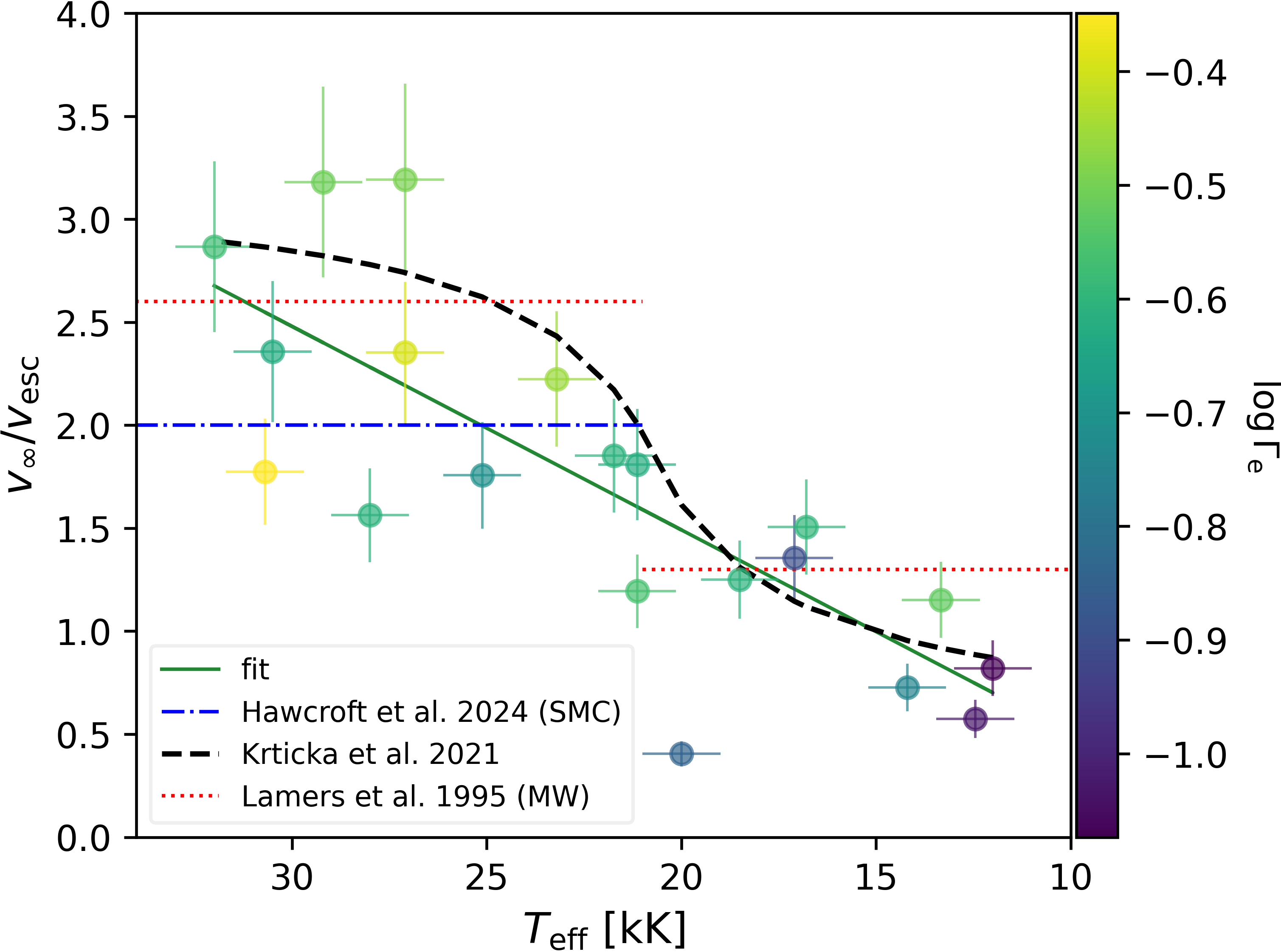}
         \caption{Ratio ($\varv_{\infty}/\varv_{{\rm esc}}$) as a function of temperature. The solid green line represents the linear fit to our sample. The dashed black line is the relation presented in \citet{krticka2021}. The dotted red lines represent the ratios from \citet{lamers1995}. The colour gradient correlates with the
         value of $\Gamma_e$.}
         \label{vinf_vesc}
    \end{figure}
In Fig.~\ref{vinf_vesc}, we present the ratio $\varv_{\infty}$ to the escape velocity $\varv_{\rm esc}$ as a function of $T_{\rm eff}$, which was calculated using $M_{\rm evo}$. From our results, we derive a relation of 
\begin{equation}
  \label{eq:vinf_vesc_t} 
  \varv_{\infty}/\varv_{\rm esc}=4.6(\pm 0.9)\log{(T_{\rm eff}/{\rm K})}-18.3(\pm 3.8),
\end{equation}
where $\varv_{\rm esc} = \sqrt{2GM(1-\Gamma_{\rm e})/R_{\ast}}$, and $\Gamma_{\rm e}$ is the Eddington parameter, which was calculated using $M_{\rm evo}$.

We do not notice any drastic decrease in $\varv_{\infty}/\varv_{\rm esc}$ below $T_{\rm eff}= 25$-$21$~kK, indicative of the theorised bi-stability jump. We find that $\varv_{\infty}/\varv_{\rm esc}$ decreases monotonically with $T_{\rm eff}$. These results show a lack of evidence for the existence of the bi-stability jump, which is similar to the findings of Paper~XIII, as well as the findings of \citet{deBurgos2024} in the MW, \citet{bernini2024} in the SMC, and \citet{verhamme2024} in the LMC. In Appendix~\ref{eta-appendix}, we include some additional discussion on the bi-stability jump.

\subsection{Implications for the population of blue supergiants}
Recalling the HRD in Fig.~\ref{HRD}, the shaded region is defined by the observed HD limit \citep{humphreys&davidson1979, humphreys&davidson1994} in the SMC, which has a value of $\log{L_{\rm bol}/L_{\odot} \approx 5.5}$ \citep{davies2018}. The diagonal blue edge of this region is inferred from the distribution of known luminous blue variables (LBVs) and LBV candidates on the HRD \citep{smith2004}.

The HD limit suggests that the most massive stars with luminosities higher than $\log{L_{\rm bol}/L_{\odot}} \approx 5.5$ remain as blue supergiants, and end their lives as blue hypergiants and classical Wolf-Rayet (WR) stars, rather than evolving into luminous red supergiants. This means that, at least in theory, they should have sufficiently high mass-loss rates to shed their hydrogen-rich atmospheres and proceed to a classical WR phase. The existence of a considerable non-binary population of WRs in the SMC \citep[e.g. ][]{hainich2015, schootemeijer2024} with high luminosity, for which \citet{shenar2020} finds a non-binary fraction of $\approx60$-$70\%$, in addition to a lower limit on WR luminosity of $\log{L_{\rm bol}/L_{\odot}} = 5.6$, which is comparable to the uppermost part of our sample and could, in principle, be seen as an argument for a self-stripping evolutionary scenario via extreme LBV eruptions.

However, aligning this evolutionary scenario with the mass-loss rates obtained in studies of blue supergiants in the SMC \citep[This study,][]{bernini2024, trundle2005, trundle2004} poses a challenge for explaining the formation of WR stars from radiation-driven mass loss alone in the single-star regime. By way of example, AzV\,242 has an evolutionary mass of $M_{\rm evo} = 37.89^{+4.01}_{-3.94}~M_{\odot}$, a luminosity of $\log{L_{\rm bol}/L_{\odot}}=5.71\pm0.1$, and a mass-loss rate of $\log{\dot{M}/M_{\odot}\,{\rm yr^{-1}}}=-7.47\pm0.4$. AzV\,242 closely corresponds to the minimum initial mass of WR stars in the SMC \citep{shenar2020}, yet the mass-loss rate of this supergiant is far too low to result in the stripping of the outer hydrogen-rich layers within its remaining $\sim$0.5 Myr lifetime. 

The low current mass-loss rates, together with the observed properties of WR populations and the lack of luminous red supergiants in the SMC, leads to the following conclusion: to form a WR star via single-star evolution, luminous blue supergiants would need to lose copious amounts of mass during an LBV giant eruption \citep{smith2017, jiang2018}. This is supported by recent work from \citet{pauli2026}, who were able to replicate the observed properties of the WR and RSG populations -- including the low-luminosity single WRs -- in the SMC by incorporating LBV-like Eddington-limit induced mass loss in evolutionary models. It is also probable that binarity, mass transfer via Roche lobe overflow, and common envelope evolution play a significant role in shaping the WR population \citep{Schootemeijer&Langer2018}. 

\section{Summary and conclusions}
We have completed a spectroscopic analysis of 20 late-O and B SMC supergiants, which employed \textsc{CMFGEN} \citep{hillier1998}, using the UV (ULLYSES) and optical (XShootU) spectral ranges. We obtained stellar and wind parameters of the stars in our sample. By comparing our results to those of \citet{alkousa2025} for the LMC, we find a clear $Z$ dependence in wind momentum, which is shown in Fig.~\ref{Dmom}. We derived the wind momentum-luminosity relation presented in Equation~\ref{eq:Dmom_Z}. We also found a clear $Z$ dependence in mass-loss rates, but did not find similar $Z$ dependence in the terminal wind velocities. 

We compared our derived mass-loss rates to predictions from various numerical recipes \cite{vinksander2021, bjorklund2023, krticka2024}. We found that the values of the numerical recipes differ from our mass-loss rates, with the recipe from \citet{krticka2024} producing the values most closely aligned with ours, yet still showing a different trend with temperature. 

We also compared our results to literature values of the same stars \citep{trundle2004, trundle2005, backs2024, bernini2024, Bestenlehner2025}. On the one hand, we find that the effective temperature, surface gravity, and luminosity generally agree with previous studies with modest variance. However, our derived mass-loss rates do not match previous estimates, with our values being generally lower than those previously obtained by previous studies.

We do not find any sign of the theorised bi-stability jump. This is similar to the findings of \citet{deBurgos2024}, \citet{bernini2024}, \citet{verhamme2024}, and \citet{alkousa2025}. This means that within the current framework of radiative transfer codes used for modelling OB stars, the dramatic increase in mass loss at $T_{\rm eff}=25$-$21~{\rm kK}$ that is associated with the theorised bi-stability jump is not empirically supported.

We also explored the evolutionary history of the stars within our sample. We employed an updated Bayesian inference method (V.~Bronner et al., in prep.) similar to \textsc{bonnsai} \citep{bonnsai2014} to obtain the evolutionary masses and ages using the rotating single-star evolutionary tracks of \citet{brott2011}. Evolutionary masses typically exceed spectroscopic masses, with $40\%$ of our sample showing a mass discrepancy. 

Finally, we emphasise that mass-loss rates of supergiants exceeding the HD limit of $\log{L_{\rm bol}/L_{\odot}} = 5.5$ \citep{davies2018} are far too low to be able to strip the supergiants of their hydrogen-rich outer layers. Consequently, the high fraction of single WR stars observed in the SMC \citep{shenar2020} cannot be explained by the classical single star evolutionary regime, unless the supergiant is capable of losing an enormous amount of mass within its very short remaining lifespan. This possibly means that an intermediate LBV phase that is accompanied by extreme episodic eruptions and mass loss is required for the star to proceed into a WR phase.

In the next installment of this series of papers, we plan to analyse a similar sample of Galactic blue supergiants using the same methods as are presented in Paper~XIII. This will allow us to derive a mass-loss rate recipe for supergiants, which avoids differences introduced by utilising various analysis techniques and codes. 

\section{Data availability}
The supplementary online material can be found on Zenodo\footnote{\href{https://doi.org/10.5281/zenodo.18416382}{https://doi.org/10.5281/zenodo.18416382}}.  \href{https://doi.org/10.5281/zenodo.18416382}{Appendix~G} includes comments on diagnostic-line fits for all the targets in our sample. 
Additionally, SED fits, diagnostic-line fits, and overall UV and optical spectral fits can be found in \href{https://doi.org/10.5281/zenodo.18416382}{Appendix~H}, \href{https://doi.org/10.5281/zenodo.18416382}{Appendix~I}, 
and \href{https://doi.org/10.5281/zenodo.18416382}{Appendix~J}, respectively.

\begin{acknowledgements}
TA would like to thank the Science and Technology Facilities Council (STFC) for financial support through the STFC scholarship ST/X508743/1. RK acknowledges financial support via the Heisenberg Research Grant funded by the Deutsche Forschungsgemeinschaft (DFG, German Research Foundation) under grant no.~KU 2849/9, project no.~445783058. F.N., acknowledges support by PID2022-137779OB-C41 funded by MCIN/AEI/10.13039/501100011033 by "ERDF A way of making Europe". AACS and MBP are supported by the German \textit{Deut\-sche
For\-schungs\-ge\-mein\-schaft, DFG\/} in the form of an Emmy Noether
Research Group -- Project-ID 445674056 (SA4064/1-1, PI Sander). This
project was co-funded by the European Union (Project 101183150 - OCEANS).This study was made possible through the Director's discretionary ULLYSES survey, which was implemented by a Space Telescope Science Institute (STScI) team led by Julia Roman-Duval. Based on observations made with ESO telescopes at the Paranal observatory under programme ID 106.211Z.001 and observations obtained with the NASA/ESA Hubble Space Telescope, retrieved from the Mikulski Archive for Space Telescopes (MAST) at the STScI. STScI is operated by the Association of Universities for Research in Astronomy, Inc. under NASA contract NAS 5-26555. We also thank John Hillier for developing CMFGEN, Nidia Morrell for obtaining and reducing MIKE data in addition to her feedback on this paper, and Vincent Bronner for facilitating the use of his Bayesian inference technique. This research has used the SIMBAD database, operated at CDS, Strasbourg, France. We would like to thank the referee for their critical and insightful comments.
\end{acknowledgements}

-------------------------------------------------------

\bibliographystyle{aa}
\bibliography{aa}
\begin{appendix} 
\section{Main results table}
\begin{sidewaystable}[!ht]
        \captionsetup{width=.85\textwidth}
        \caption{Derived stellar parameters based on the best fitting \textsc{CMFGEN} model.}            
        \label{table:2}      
        \def\arraystretch{1.8}
        \small
        \addtolength{\tabcolsep}{-0.2em}
        \begin{tabular}{c c c c c | c  c c c c | c c c | c c c c}  
            \hline\hline{\smallskip}
Star&$T_{\rm eff}$ &Diag.&$\log{g_c}$ &$R_{*}$ &$\log{L_{\rm bol}}$&$A_{V}$ &$M_{\rm V}$ &$BC_{\rm V}$ &$v_{\rm rad}$ &$v_{\rm rot}\sin{i}$&$M_{\rm spec}$&$\Gamma_{\rm e}^{\rm spec}$ &$M_{init}$&$M_{\rm evo}$&Age&$\Gamma_{\rm e}^{\rm evo}$\\
&kK&&${\rm cm\,s^{-2}}$&$R_{\odot}$&$L_{\odot}$&mag&mag&mag&${\rm km\,s^{-1}}$&${\rm km\,s^{-1}}$&$M_{\odot}$&&$M_{\odot}$&$M_{\odot}$&Myr&\\
            \hline
AzV\,469      &$32.0$&$\ion{He}{I}-\ion{He}{II}$&$3.30$&$19\pm1$  &$5.55$&$0.27$&$-6.11$&$-3.03$&$185$ &$55$&$27.6\pm3.9$ &$0.29\pm0.06$ &$32.24^{+7.60}_{-6.68}$&$29.89^{+9.21}_{-6.12}$ &$4.34^{+1.52}_{-0.63}$&$0.27^{+0.08}_{-0.06}$\\
AzV\,372      &$29.2$&$\ion{He}{I}-\ion{He}{II}$&$3.00$&$27\pm2$  &$5.68$&$0.28$&$-6.65$&$-2.80$&$251$ &$100$&$28.1\pm4.3$&$0.38\pm0.08$ &$37.20^{+5.74}_{-5.88}$&$34.73^{+7.15}_{-5.52}$ &$4.86^{+0.32}_{-1.11}$&$0.31^{+0.06}_{-0.05}$\\
AzV\,456      &$30.7$&$\ion{He}{I}-\ion{He}{II}$&$3.35$&$32\pm2$  &$5.91$&$0.97$&$-7.10$&$-2.94$&$141$ &$55$&$83.4\pm11.8$&$0.25\pm0.05$ &$50.34^{+5.75}_{-5.29}$&$46.77^{+6.39}_{-3.65}$ &$3.26^{+0.21}_{-0.33}$&$0.45^{+0.06}_{-0.04}$\\
AzV\,327      &$30.5$&$\ion{He}{I}-\ion{He}{II}$&$3.32$&$19\pm1$  &$5.47$&$0.09$&$-6.02$&$-2.92$&$179$ &$55$&$29.1\pm4.1$ &$0.24\pm0.05$ &$30.14^{+3.78}_{-5.13}$&$29.47^{+3.35}_{-5.16}$ &$4.90^{+0.65}_{-0.76}$&$0.23^{+0.03}_{-0.04}$\\
AzV\,235      &$27.1$&$\ion{Si}{III}-\ion{Si}{IV}$&$2.98$&$39\pm3$  &$5.86$&$0.52$&$-7.28$&$-2.64$&$165$ &$60$&$52.7\pm8.4$ &$0.35\pm0.08$ &$48.08^{+5.20}_{-6.51}$&$46.13^{+4.22}_{-6.48}$ &$3.46^{+0.38}_{-0.30}$&$0.40^{+0.04}_{-0.06}$\\
AzV\,215      &$27.1$&$\ion{Si}{III}-\ion{Si}{IV}$&$2.95$&$29\pm2$  &$5.62$&$0.42$&$-6.69$&$-2.63$&$165$ &$60$&$28.4\pm4.5$ &$0.38\pm0.08$ &$34.74^{+4.80}_{-3.70}$&$33.44^{+4.75}_{-3.16}$ &$4.34^{+0.52}_{-0.47}$&$0.32^{+0.05}_{-0.03}$\\
AzV\,104      &$28.0$&$\ion{Si}{III}-\ion{Si}{IV}$&$3.25$&$22\pm2$  &$5.42$&$0.28$&$-6.07$&$-2.72$&$140$ &$65$&$31.2\pm4.6$ &$0.22\pm0.05$ &$27.96^{+3.08}_{-2.71}$&$27.48^{+2.80}_{-2.71}$ &$5.22^{+0.57}_{-0.49}$&$0.25^{+0.03}_{-0.03}$\\
AzV\,410      &$25.1$&$\ion{Si}{III}-\ion{Si}{IV}$&$3.10$&$23\pm2$  &$5.26$&$0.19$&$-5.94$&$-2.47$&$185$ &$85$&$24.2\pm3.8$ &$0.19\pm0.04$ &$23.28^{+2.45}_{-2.22}$&$23.17^{+2.15}_{-2.43}$ &$6.43^{+0.89}_{-0.67}$&$0.20^{+0.02}_{-0.02}$\\
AzV\,242      &$23.2$&$\ion{Si}{III}-\ion{Si}{IV}$&$2.65$&$44\pm4$  &$5.71$&$0.35$&$-7.25$&$-2.25$&$177$ &$55$&$32.7\pm5.9$ &$0.40\pm0.09$ &$39.47^{+4.69}_{-4.18}$&$37.89^{+4.01}_{-3.94}$ &$4.15^{+0.41}_{-0.34}$&$0.35^{+0.04}_{-0.04}$\\
AzV\,96       &$21.7$&$\ion{Si}{III}-\ion{Si}{IV}$&$2.65$&$36\pm3$  &$5.41$&$0.30$&$-6.67$&$-2.12$&$158$ &$50$&$21.3\pm3.9$ &$0.31\pm0.07$ &$27.78^{+2.21}_{-3.74}$&$26.88^{+2.14}_{-3.35}$ &$5.87^{+0.64}_{-0.69}$&$0.25^{+0.02}_{-0.03}$\\
AzV\,264      &$21.1$&$\ion{Si}{III}-\ion{Si}{IV}$&$2.50$&$39\pm4$  &$5.44$&$0.23$&$-6.82$&$-2.04$&$131$ &$45$&$18.1\pm3.5$ &$0.36\pm0.08$ &$28.40^{+3.85}_{-4.17}$&$27.31^{+3.64}_{-3.68}$ &$5.89^{+0.58}_{-1.02}$&$0.24^{+0.03}_{-0.03}$\\
AzV\,175      &$20.0$&$\ion{Si}{III}-\ion{Si}{IV}$&$2.80$&$25\pm3$  &$4.97$&$0.31$&$-5.74$&$-1.93$&$175$ &$30$&$15.0\pm2.7$ &$0.16\pm0.04$ &$17.11^{+1.80}_{-1.31}$&$16.69^{+2.07}_{-1.01}$ &$9.30^{+1.22}_{-0.99}$&$0.14^{+0.02}_{-0.01}$\\
Sk 191       &$21.1$&$\ion{Si}{III}-\ion{Si}{IV}$&$2.45$&$51\pm5$  &$5.67$&$0.52$&$-7.62$&$-2.06$&$127$ &$65$&$27.9\pm5.4$ &$0.4\pm0.09$ &$43.04^{+3.56}_{-3.14}$&$40.81^{+3.13}_{-2.93}$ &$3.98^{+0.28}_{-0.23}$&$0.27^{+0.02}_{-0.02}$\\
AzV\,18       &$18.5$&$\ion{Si}{II}-\ion{Si}{III}$&$2.25$&$52\pm6$  &$5.45$&$0.66$&$-7.16$&$-1.73$&$150$ &$50$&$18.0\pm3.9$ &$0.37\pm0.09$ &$27.50^{+3.67}_{-3.03}$&$27.21^{+2.88}_{-3.47}$ &$5.57^{+0.80}_{-0.62}$&$0.25^{+0.03}_{-0.03}$\\
NGC330-ELS-04&$17.1$&$\ion{Si}{II}-\ion{Si}{III}$&$2.35$&$33\pm4$  &$4.93$&$0.37$&$-6.00$&$-1.57$&$158$ &$35$&$9.2\pm2.0$  &$0.46\pm0.10$ &$16.58^{+1.64}_{-1.31}$&$16.49^{+1.59}_{-1.32}$ &$9.79^{+1.26}_{-1.10}$&$0.13^{+0.01}_{-0.01}$\\
AzV\,187      &$16.8$&$\ion{Si}{II}-\ion{Si}{III}$&$2.20$&$57\pm7$  &$5.37$&$0.25$&$-7.15$&$-1.54$&$139$ &$40$&$19.4\pm4.4$ &$0.22\pm0.05$ &$24.51^{+4.07}_{-3.72}$&$24.01^{+4.41}_{-4.26}$ &$6.00^{+1.38}_{-0.80}$&$0.25^{+0.05}_{-0.05}$\\
AzV\,22       &$14.2$&$\ion{Si}{II}-\ion{Si}{III}$&$1.80$&$67\pm10$ &$5.22$&$0.38$&$-7.14$&$-1.17$&$145$ &$30$&$10.7\pm2.9$ &$0.31\pm0.08$ &$21.94^{+2.18}_{-2.33}$&$21.65^{+2.06}_{-2.22}$ &$7.18^{+0.87}_{-0.97}$&$0.18^{+0.02}_{-0.02}$\\
AzV\,393      &$13.3$&$\ion{He}{I}-\ion{Mg}{II}$&$1.60$&$109\pm16$&$5.53$&$0.48$&$-8.01$&$-1.09$&$151$ &$30$&$17.8\pm5.3$ &$0.49\pm0.13$ &$30.41^{+3.68}_{-4.02}$&$29.23^{+8.51}_{-10.12}$&$5.16^{+0.76}_{-0.60}$&$0.30^{+0.09}_{-0.11}$\\
AzV\,343      &$12.0$&$\ion{He}{I}-\ion{Mg}{II}$&$1.98$&$49\pm8$  &$4.65$&$0.31$&$-6.14$&$-0.75$&$160$ &$35$&$8.6\pm2.4$  &$0.13\pm0.03$ &$13.21^{+1.12}_{-1.07}$&$12.90^{+1.32}_{-0.81}$ &$13.30^{+1.99}_{-1.55}$&$0.08^{+0.01}_{-0.01}$\\
AzV\,324      &$12.5$&$\ion{He}{I}-\ion{Mg}{II}$&$2.20$&$48\pm8$  &$4.70$&$0.06$&$-6.18$&$-0.83$&$156$ &$25$&$13.5\pm3.5$ &$0.09\pm0.03$ &$13.73^{+1.17}_{-1.07}$&$13.56^{+1.24}_{-1.04}$ &$12.54^{+1.79}_{-1.46}$&$0.09^{+0.01}_{-0.01}$\\
            \noalign{\smallskip}
            \hline            
            \FloatBarrier
        \end{tabular}
  \vspace{9cm}
\end{sidewaystable}
\clearpage
In Table~\ref{table:2}, $M_{init}$, $M_{\rm evo}$, and ages of the stars are derived using an updated Bayesian inference method (Bronner et al. in prep) that is similar to {\sc Bonnsai} \citep{bonnsai2014} applied to \citet{brott2011} SMC evolutionary tracks. The average uncertainty in $T_{\rm eff}$ is $\Delta T_{\rm eff}=1~{\rm kK}$. In the uncertainties of centrifugal force-corrected surface gravity $\log{g_c}$, we take into account the fitting uncertainty of $\log{g}$ and the uncertainty of the projected rotational velocity $v_{\rm rot}\sin{i}$ and the radius $R_{*}$. We find a mean uncertainty of $\Delta \log{g_c} = 0.2$~dex. The uncertainty in radial velocity $v_{\rm rad}$ measurements is dominated by the velocity resolution of the UBV part of the spectrum which is $\Delta\varv\approx45~{\rm km\,s^{-1}}$. The total extinction $A_{V}$ ($\Delta A_{V} \approx 0.02$~mag), absolute V-band magnitude $M_{\rm V}$, and bolometric correction $BC_{\rm V}$ are produced from the SED fits. $\Gamma_{\rm e}^{\rm spec}$ is subject to uncertainties of $\Delta\Gamma_{\rm e}^{\rm spec} \approx 0.2$ which takes into account the uncertainties of $M_{\rm spec}$ and $\log{L_{\rm bol}}$

\section{Additional discussion on $Z$ dependence}
\label{Verhamme}
In parallel with the present study, \citet{verhamme2025} have analysed a sample of ULLYSES/XShootU SMC B supergiants using \textsc{KIWI-GA}. They find broadly comparable mass-loss rates and wind momenta with respect to LMC B supergiants \citep{verhamme2024}, in contrast with our results, despite having 13 stars in common with our sample. 

This surprising result could be attributed to a number of issues. Firstly, the  determination of robust wind parameters of late B supergiants in the SMC is particularly challenging due to the lack of wind diagnostics, and the current version of \textsc{FASTWIND} is known to have convergence difficulties below 14~kK \citep{deburgos2024b}. These issues could have skewed \textsc{KIWI-GA} solutions to higher $D_{\rm mom}$. This is confirmed when examining the H$\alpha$ fits of \citet{verhamme2025}, where the preferred solution overpredicts H$\alpha$ emission. In addition, their derived $T_{\rm eff}$ often deviates quite significantly from the usual spectral type -- temperature scale of B supergiants \citep[see e.g. ][]{lanzandhubeny2007, trundle2007, nieva2013}. By way of example, \citet{verhamme2025} obtained a lower $T_{\rm eff}$ for AzV\,410 (B0.7 Iab) than NGC330\,ELS\,4 (B2.5 Ib).

\section{Additional discussion of wind parameters and the bi-stability jump}\label{eta-appendix}
\label{app:bistab}
\begin{figure}[!htbp]
    \centering
     \includegraphics[width=\hsize]{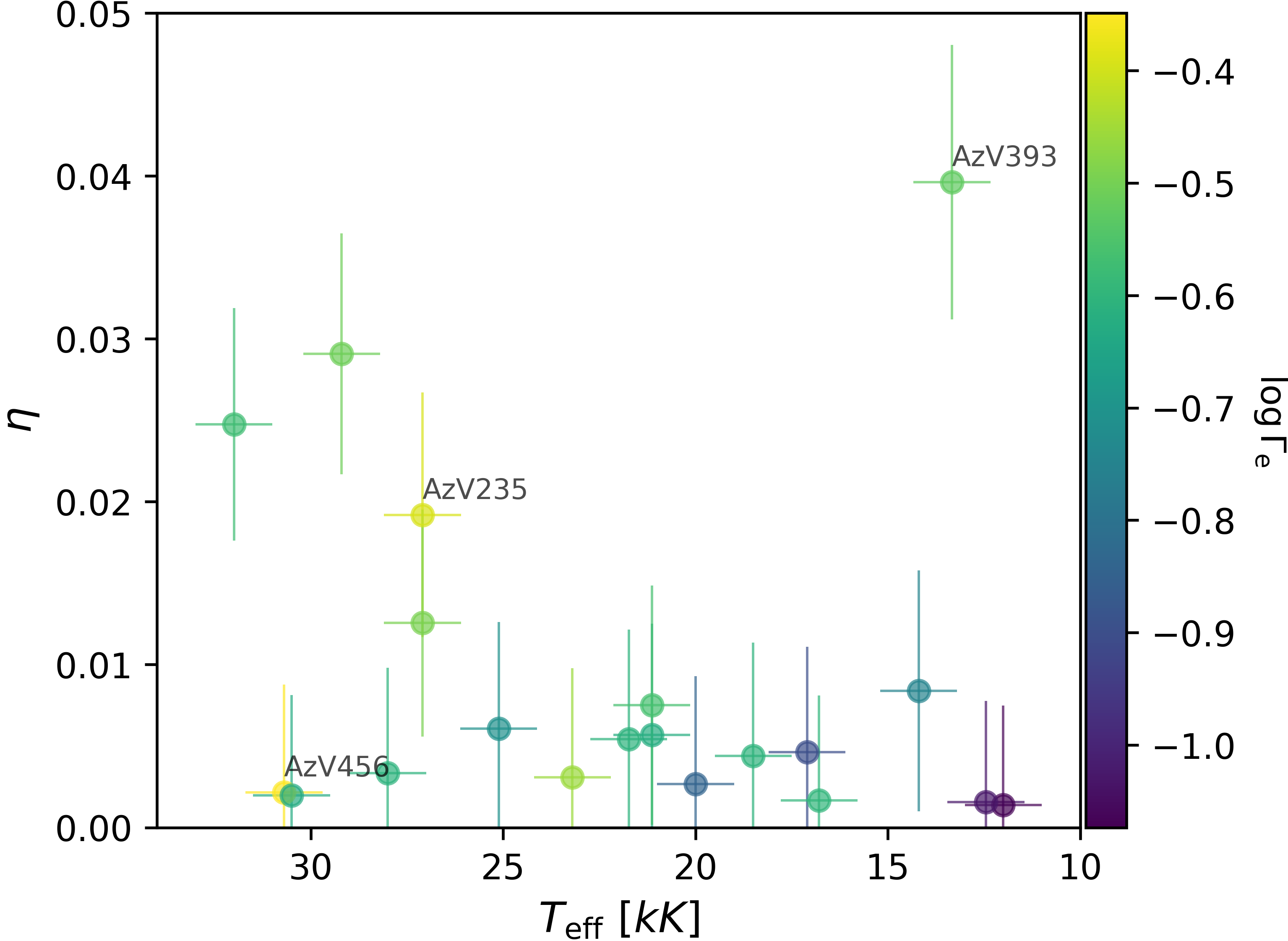}
         \caption{Wind efficiency $\eta = \frac{\dot{M}\varv_{\infty}}{L_{\rm bol}/c}$ in terms of $T_{\rm eff}$.}
         \label{eta}
         \FloatBarrier
    \end{figure}
In Fig.~\ref{eta}, we show the wind efficiency parameter $\eta = \frac{\dot{M}\varv_{\infty}}{L_{\rm bol}/c}$ versus $T_{\rm eff}$. We find that $\eta$ smoothly decreases with $T_{\rm eff}$. One extreme outlier is the hypergiant AzV\,393, which has a very high $\dot{M}$. Another outlier is AzV\,456, which has a very low $\eta$. This is due to its $\varv_{\infty}$, which is lower by $\approx300~{\rm km\,s^{-1}}$ compared to AzV\,327, which has a very similar $T_{\rm eff}$.

\section{CNO diagnostics}
\label{app:atomic}
Table~\ref{table:CNO} shows the diagnostic lines used to constrain the surface abundances of carbon, nitrogen, and oxygen. It also shows the spectral type range in which each metal line is used. 

\begin{table}[!h]
        \caption{Line diagnostics for determination of CNO elemental abundances.}
        \label{table:CNO} 
        \centering            
        \small
        \begin{tabular}{c c}        
        \hline
        line    &Sp. T.\\
        \hline
         $\ion{C}{IV}~\lambda\lambda5801-5811$ &O9-B0\\
         $\ion{C}{III}~\lambda\lambda4647-4650$ &O9-B3\\
         $\ion{C}{III}~\lambda5696$ &O9-B3\\
         $\ion{C}{II}~\lambda4070$ &B0-B8\\
         $\ion{C}{II}~\lambda4267$ &B0-B8\\
         $\ion{C}{II}~\lambda\lambda6578-6582$ &B0-B8\\
     
         $\ion{N}{III}~\lambda4097$ &O9-B0.5\\
         $\ion{N}{III}~\lambda\lambda4379$ &O9-B0.5\\
         $\ion{N}{III}~\lambda\lambda4510-4515$ &O9-B0\\
         $\ion{N}{III}~\lambda\lambda4634-4641$ &O9-B0\\
         $\ion{N}{II}\,\lambda3995$ &B1-B8\\
         $\ion{N}{II}\,\lambda4447$ &B1-B8\\
         $\ion{N}{II}\,\lambda\lambda\lambda4601-4607-4614$ &B1-B8\\
         $\ion{N}{II}\,\lambda4630$ &B1-B8\\

         $\ion{O}{III}~\lambda\lambda3261-3265$ &O9-B1\\
         $\ion{O}{III}~\lambda3760$ &O9-B1\\
         $\ion{O}{III}~\lambda5592$ &O9-B1\\
         $\ion{O}{II}~\lambda4254$ &B0-B6\\
         $\ion{O}{II}~\lambda4367$ &B0-B6\\
         $\ion{O}{II}~\lambda\lambda4415-4417$ &B0-B6\\
         $\ion{O}{II}~\lambda\lambda4638-4641$ &B0-B6\\
         $\ion{O}{I}~\lambda\lambda\lambda7772-7774-7775$ &B7-B8\\
         \hline
        \end{tabular}
    \end{table}

\section{$\varv_{\rm black}$ and $\varv_{\rm edge}$}
\label{app:vinf}
In Table~\ref{table_app_2} we present the measured $\varv_{\rm edge}$ and $\varv_{\rm black}$ from each line for all our sample.

    \begin{table}
      \begin{threeparttable}[H]
        \caption{$\varv_{\rm edge}$ and $\varv_{\rm black}$  for the stars in our sample.}
        \def\arraystretch{1}     
        \label{table_app_2}     
        \centering                                     
        \small                                
        \addtolength{\tabcolsep}{-0.3em}
        \begin{tabular}{c c c c c c c}  
            \hline
            \multicolumn{7}{c}{$\varv_{\rm edge}$}\\
            \hline
            Target &$\ion{Si}{iv}$ &$\ion{Si}{iv}$ &$\ion{C}{iv}$ &$\ion{Al}{III}$ &$\ion{Al}{III}$ &$\overline{\varv_{\rm black}}/\overline{\varv_{\rm edge}}$ \\
            &$1393{\rm \AA}$ &$1403{\rm \AA}$ &$1548{\rm \AA}$ &$1855{\rm \AA}$ &$1863{\rm \AA}$ & \\
                 
            \hline
        AzV\,469        &$2190$ &- &$2133$ &- &- &$0.84$\\
        AzV\,372        &$1944$ &- &- &- &- &$0.78$\\
        AzV\,456        &$1655$ &- &$2327$ &- &- &-\\
        AzV\,327        &$1935$ &- &$1976$ &- &- &-\\
        AzV\,235        &$1653$ &- &$1633$ &- &- &-\\
        AzV\,215        &$1935$ &- &$2060$ &- &- &-\\
        AzV\,104        &- &- &$1255$      &- &- &-\\
        AzV\,410        &$1218$ &- &$1280$ &- &- &-\\
        AzV\,242        &$1105$ &- &$1270$ &- &- &-\\
        AzV\,96         &$972$  &- &$956$  &- &- &-\\
        AzV\,264        &$827$ &- &- &- &- &-\\
        AzV\,175        &$222$ &- &$224$ &- &- &-\\
        Sk 191          &$620$ &$594$ &$600$ &$582$ &- &$0.76$\\
        AzV\,18         &$507$ &- &$472$ &- &- &-\\
        ELS-04          &$515$ &- &- &- &- &-\\
        AzV\,187        &$589$ &$588$ &$575$ &- &- &-\\
        AzV\,22         &$190$ &$173$ &$206$ &$222$ &$230$ &-\\
        AzV\,393        &$319$ &- &$279$ &- &- &-\\
        AzV\,343        &$285$ &$251$ &$282$ &- &- &-\\
        AzV\,324        &$259$ &$207$ &$220$ &$210$ &- &-\\ 
            \noalign{\smallskip}
            \hline
            \multicolumn{7}{c}{$\varv_{\rm black}$}\\
            \hline
            Target &$\ion{Si}{iv}$ &$\ion{Si}{iv}$ &$\ion{C}{iv}$ &$\ion{Al}{III}$ &$\ion{Al}{III}$ &$\overline{\varv_{\rm black}}$ \\
            &$1393{\rm \AA}$ &$1403{\rm \AA}$ &$1548{\rm \AA}$ &$1855{\rm \AA}$ &$1863{\rm \AA}$ &${\rm km\,s^{-1}}$ \\
            \hline
        AzV\,469        &- &- &$1814$ &- &- &$1814\pm100$\\
        AzV\,372        &- &- &$1516$ &- &- &$1516\pm100$\\
        Sk 191          &$474$ &$440$ &$451$ &- &- &$455\pm45$\\
            \noalign{\smallskip}
            \hline
        \end{tabular}
        \vspace{-0.3cm}
        \tablefoot{We obtain $\varv_{\rm edge}$ and $\varv_{\infty}$ from all the available P Cygni line profiles in the UV. Single velocity measurement from each line is subject to an uncertainty $\Delta\varv = 30~{\rm km\,s^{-1}}$ which takes into account the velocity resolution of the UV spectra and the uncertainty in the radial velocity correction, which were added in 
        quadrature.}
      \end{threeparttable}
    \end{table}
\FloatBarrier

\section{Comparison to literature}
\label{app:comp}
Table~\ref{comparison_table} contains our derived values for $T_{\rm eff}$, $\log{g}$, $L_{\rm bol}$, $\log{\dot{M}}$, $\beta$, and $f_{\rm vol,\infty}$, along with literature values from \citet{trundle2004}, \citet{trundle2005}, \citet{bernini2024}, \citet{backs2024}, and \citet{Bestenlehner2025}. The volume-filling factor for studies that used the optically thick clumping implementation in \textsc{FASTWIND} is calculated from the equation introduced in \citet{sander2024} following: 
\begin{equation}
    \label{eq:BC} 
    f_{\rm vol,\infty} = \frac{(1-f_{\rm ic})^2}{f_{\rm cl}-2f_{\rm ic}+f_{\rm ic}^2},
\end{equation}
where $f_{\rm cl} = \langle\rho^2\rangle/\langle\rho\rangle^2$ describes the contrast between the density in the clump and the mean density, and $f_{\rm ic}$ describes the contrast between the clump density and the density of the inter-clump medium.

\begin{figure*}
    \centering
     \includegraphics[width=\hsize]{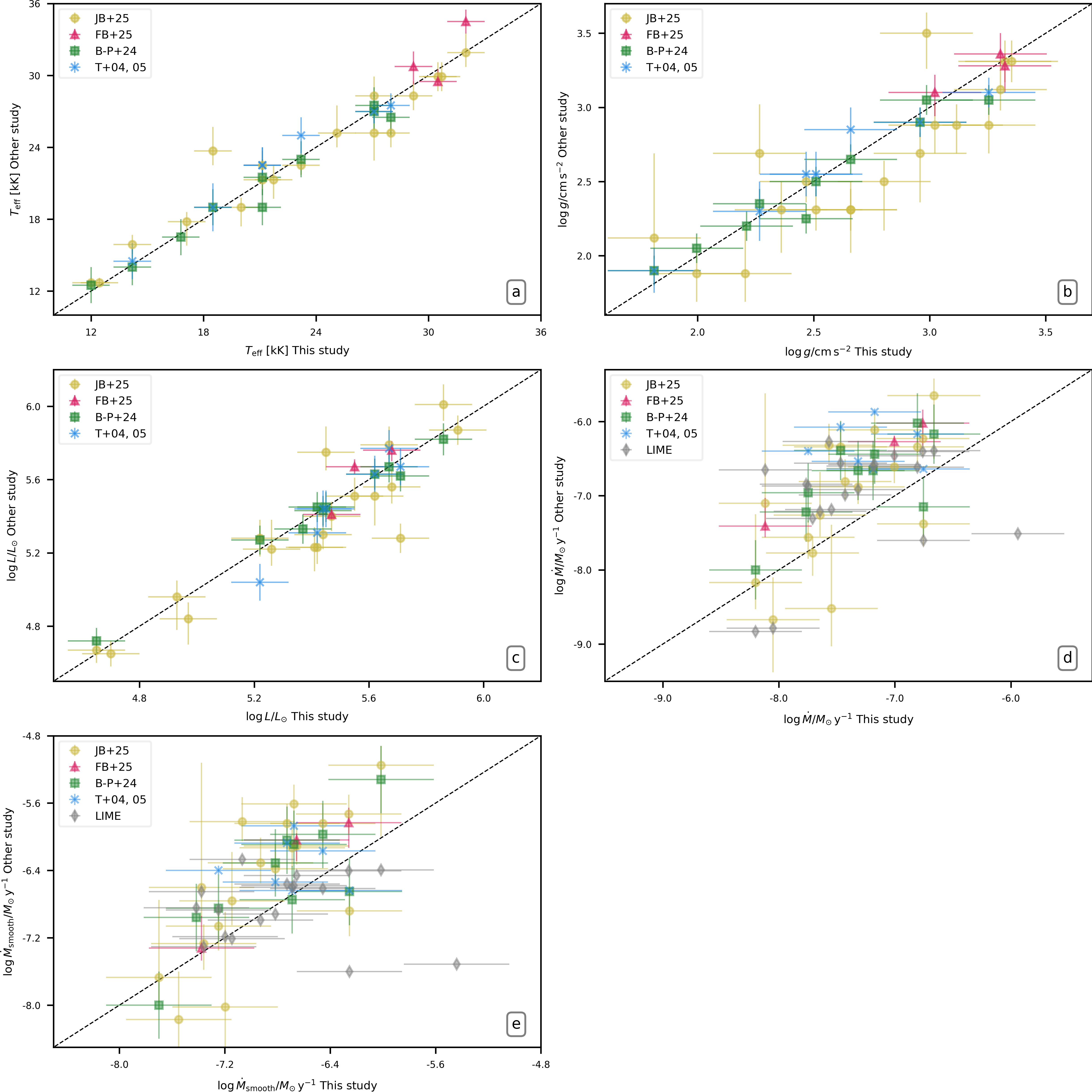}
         \caption{Comparison of our obtained parameters to literature values. a: $T_{\rm eff}$ comparison, b: $\log{g}$ comparison, c: $L_{\rm bol}$ comparison, d: $\log{\dot{M}}$ comparison, e: $\log{\dot{M}_{\rm smooth}}$ comparison. Pink triangles: comparison to \citet{backs2024}. Green squares: comparison to \citet{bernini2024}. Blue $X$s: comparison to \citet{trundle2004} and \citet{trundle2005}. Grey diamonds: comparison to values obtained from LIME \citet{LIME}.}
         \label{comp_lit}
    \end{figure*}

In addition to literature values, Table~\ref{comparison_table} in includes $\log{\dot{M}}$ values obtained from \textsc{LIME} \citep{LIME} . \textsc{LIME} is a user-friendly and fast online tool that allows for the estimation of the mass-loss rates of stars using an iterative semi-analytic approach using an extension of the \cite{CAK} concept. By including a detailed list of atomic lines, but avoiding a numerical solution for the equation of motion and the stellar atmosphere, the tool can estimate $\dot{M}$ on the fly, in contrast to the recipes discussed in Section~\ref{subsec:wind_properties}, which result from approximating sets of detailed numerical models. The input parameters we used for LIME are our derived $T_{\rm eff}$, $L_{\rm bol}$, $M_{\rm spec}$, mass fractions of H, He, C, N, O, and the mean mass fractions for elements heavier than oxygen presented in \citet{xshootu1}.                     
In Fig.~\ref{comp_lit} we compare our obtained values $T_{\rm eff}$, $\log{g}$, $L_{\rm bol}$, $\log{\dot{M}}$, and $\log{\dot{M}_{\rm smooth}} = \log{\dot{M}/\sqrt{f_{\rm vol,\infty}}}$ to the values in the literature. We find that our $T_{\rm eff}$ values are consistent with values from the literature within the uncertainties. The only exception is AzV\,18 for which \citet{Bestenlehner2025} find a value of $T_{\rm eff}=23.7$~kK compared to our value of $18.5$~kK. Similarly, our $\log{g}$ is generally in agreement with values from the aforementioned studies. We find that $\log{g}$ values obtained by \citet{Bestenlehner2025} are systematically lower than ours.

We find a similar picture for $L_{\rm bol}$, where our derived values are consistent with the values presented in the literature. The exception in this case is some of the values obtained by \citet{Bestenlehner2025}, mainly for AzV\,235, AzV\,242, and AzV\,18, for which we obtain $\log{L_{\rm bol}/L_{\odot}} =5.86,\, 5.71,\,5.45$, respectively, compared to $6.01,\, 5.28,\, 5.75$ from \citet{Bestenlehner2025}.

The $\dot{M}$ comparison tells a different story. We find that for the most part, the values of $\dot{M}$ are inconsistent, and our $\dot{M}$ is generally lower. The mean difference between our $\log{\dot{M}}$ and those from \citet{Bestenlehner2025}, \citet{backs2024}, \citet{bernini2024}, \citet{trundle2004} and \citet{trundle2005}, and \textsc{LIME} are $\approx -0.4, -0.7, -0.5, -0.9, -0.3$~dex, respectively.
The immediate contributors to this inconsistency are the choice of clumping implementation (optically-thin vs optically-thick clumping and clumping law), the derived values for clumping parameters, and the derived/adopted values of $\beta$ and $\varv_{\infty}$. Furthermore, the differences in $T_{\rm eff}$, $\log{g}$, and $\epsilon_{\rm He}$ propagate to the estimated $\dot{M}$. Another obvious issue is the usage of various codes that implement different methods for solving the radiative transfer equations. Therefore, when all these issues are taken into account, differences up to a factor of 3 in $\dot{M}$ can be expected. 

We can test if the values of clumping parameters are a major cause of the inconsistency in the mass-loss rates by comparing the `smooth' mass-loss rates, $\dot{M}_{\rm smooth}$. This shows a similar trend; our $\dot{M}_{\rm smooth}$ is systematically lower than other studies, except for the values obtained from \textsc{LIME}, but now the mean difference between our $\dot{M}_{\rm smooth}$  and those from the literature are considerably lower and are $\approx -0.4, -0.4, -0.3, -0.4, 0.2$~dex, respectively. Therefore, the derived/adopted values of the clumping parameters can significantly affect the resulting mass loss rates and are a major factor in the discrepancy seen in $\dot{M}$.

\begin{table*}[!h]
  \centering   
\begin{threeparttable}[b]
    \caption{Comparison of derived parameters to previous studies.}              
        \def\arraystretch{1.2}
        \label{comparison_table}                             
        \small                               
        \addtolength{\tabcolsep}{0.5em}
        \begin{tabular}{c c c c c c c c c c}          
            \hline
            Target  &$T_{\rm eff}$          &$\log{g}$              &$\log{L}$              &$\log{\dot{M}}$            &$f_{\rm vol,\infty}$  &$\beta$  &Ref.       &wavelength &Code \\[1.50 pt]
            
                    &kK                     &${\rm cm\,s^{-2}}$     &$L_{\odot}$            &$M_{\odot}\,{\rm yr}^{-1}$ &   &       &           &       &\\     
            \hline
            AzV 469 &$32.0$                 &$3.30$                 &$5.55$                 &$-7.00$                &$0.2$  &$1.0$  &This study &UV+OPT     &C\\
                    &$31.9^{+1.2}_{-1.6}$    &$3.12^{+0.1}_{-0.2}$    &$5.51^{+0.1}_{-0.1}$    &$-6.61^{+0.2}_{-0.2}$   &$0.1$  &$1.0$  &JB+25      &OPT        &F\\
                    &$34.5^{+1.0}_{-1.0}$    &$3.36^{+0.1}_{-0.1}$    &$5.67^{+0.0}_{-0.0}$    &$-6.27^{+0.2}_{-0.1}$   &$0.34$ &$1.1$  &FB+24      &UV+OPT     &F\\
                    &-                      &-                      &-                      &$-6.46$                &-      &-      &LIME       &-          &-\\
                    
            AzV 372 &$29.2$                 &$3.02$                 &$5.68$                 &$-6.76$                &$0.1$  &$1.0$  &This study &UV+OPT     &C\\
                    &$28.3^{+1.2}_{-1.2}$    &$2.88^{+0.2}_{-0.1}$    &$5.56^{+0.1}_{-0.1}$    &$-6.23^{+0.2}_{-0.2}$   &$0.1$  &$1.0$  &JB+25      &OPT        &F\\
                    &$30.8^{+1.5}_{-1.2}$    &$3.1^{+0.2}_{-0.1}$     &$5.76^{+0.1}_{-0.0}$    &$-6.02^{+0.3}_{-0.2}$   &$0.42$ &$1.8$  &FB+24      &UV+OPT     &F\\
                    &-                      &-                      &-                      &$-6.4$                 &-      &-      &LIME       &-          &-\\
                    
            AzV 456 &$30.7$                 &$3.35$                 &$5.91$                 &$-7.57$                &$0.1$  &$1.0$  &This study &UV+OPT     &C\\
                    &$29.9^{+1.2}_{-1.2}$    &$3.31^{+0.1}_{-0.1}$    &$5.87^{+0.1}_{-0.1}$    &$-6.32^{+0.2}_{-0.3}$   &$0.1$  &$1.0$  &JB+25      &OPT        &F\\
                    &-                      &-                      &-                      &$-6.27$                &-      &-      &LIME       &-          &-\\
            
            AzV 327 &$30.5$                  &$3.32$                  &$5.47$                 &$-8.12$                &$0.03$ &$1.0$  &This study &UV+OPT     &C\\
                    &$29.9^{+1.2}_{-1.2}$    &$3.31^{+0.1}_{-0.1}$    &$5.40^{+0.1}_{-0.1}$   &$-7.1^{+0.4}_{-1.5}$   &$0.1$  &$1.0$  &JB+25      &OPT        &F\\
                    &$29.5^{+0.2}_{-1.0}$    &$3.28^{+0.2}_{-0.1}$    &$5.41^{+0.0}_{-0.0}$   &$-7.41^{+0.2}_{-0.4}$  &$0.67$ &$0.8$  &FB+24      &UV+OPT     &F\\
                    &-                       &-                       &-                      &$-6.65$                &-      &-      &LIME       &-          &-\\
            
            AzV 235   &$27.1$                 &$2.99$                 &$5.86$                 &$-6.66$                &$0.05$ &$1.5$  &This study &UV+OPT     &C\\
                    &$28.3^{+1.2}_{-1.6}$    &$3.5^{+0.2}_{-0.1}$     &$6.01^{+0.1}_{-0.1}$    &$-5.65^{+0.8}_{-0.2}$   &$0.1$  &$1.0$  &JB+25      &OPT        &F\\
                    &$27.5$                 &$3.05$                 &$5.82$                 &$-6.17$                &$0.02$ &$2.4$  &BP+24      &UV+OPT     &C\\
                    &-                      &-                      &-                      &$-6.39$                &-      &-      &LIME       &-          &-\\
            
            AzV 215 &$27.1$                 &$2.96$                 &$5.62$                 &$-7.17$                &$0.1$  &$1.7$  &This study &UV+OPT     &C\\
                    &$25.2^{+2.3}_{-1.2}$    &$2.69^{+0.3}_{-0.1}$    &$5.51^{+0.2}_{-0.1}$    &$-6.11^{+0.2}_{-0.2}$   &$0.1$  &$1.0$  &JB+25      &OPT        &F\\
                    &$27.0$                 &$2.9$                  &$5.63$                 &$-6.44$                &$0.2$  &$2.8$  &BP+24      &UV+OPT     &C\\
                    &$27.0^{+1.0}_{-1.0}$    &$2.9^{+0.1}_{-0.1}$     &$5.63$                 &$-5.87$                &$1$    &$1.4$  &T+04       &OPT        &F\\
                    &-                      &-                      &-                      &$-6.58$                &-      &-      &LIME       &-          &-\\
                    
            AzV 104 &$28.0$                 &$3.25$                 &$5.42$                 &$-7.75$                &$0.1$  &$1.0$  &This study &UV+OPT     &C\\
                    &$25.2^{+1.2}_{-1.2}$    &$2.88^{+0.2}_{-0.1}$    &$5.23^{+0.1}_{-0.1}$    &$-7.56^{+0.3}_{-0.2}$   &$0.1$  &$1.0$  &JB+25      &OPT        &F\\
                    &$26.5$                 &$3.05$                 &$5.45$                 &$-6.96$                &$0.6$  &$0.5$  &BP+24      &UV+OPT     &C\\
                    &$27.5^{1.0}_{-1.0}$    &$3.1^{0.1}_{-0.1}$     &$5.31$                 &$-6.4$                 &$1$    &$1.0$  &T+04       &OPT        &F\\
                    &-                      &-                      &-                      &$-6.87$                &-      &-      &LIME       &-          &-\\

            AzV 410 &$25.1$                 &$3.12$                 &$5.26$                 &$-7.65$                &$0.1$  &$1.0$  &This study &UV+OPT     &C\\
                    &$25.2^{+1.2}_{-2.3}$    &$2.88^{+0.2}_{-0.1}$    &$5.22^{+0.1}_{-0.2}$    &$-7.26^{+0.3}_{-0.6}$   &$0.1$  &$1.0$  &JB+25      &OPT        &F\\
                    &-                      &-                      &-                      &$-7.21$                &-      &-      &LIME       &-          &-\\
            
            AzV 242 &$23.2$                 &$2.66$                 &$5.71$                 &$-7.47$                &$0.03$ &$3.1$  &This study &UV+OPT     &C\\
                    &$22.5^{+0.8}_{-0.8}$    &$2.31^{+0.1}_{-0.1}$    &$5.28^{+0.1}_{-0.1}$    &$-6.34^{+0.2}_{-0.2}$   &$0.1$  &$1.0$  &JB+25      &OPT        &F\\
                    &$23.0$                 &$2.65$                 &$5.62$                 &$-6.39$                &$0.2$  &$1.8$  &BP+24      &UV+OPT     &C\\
                    &$25.0^{+1.5}_{-1.5}$    &$2.85^{+0.2}_{-0.2}$    &$5.67$                 &$-6.08$                &$1$    &$2.0$  &T+05       &OPT        &F\\
                    &-                      &-                      &-                      &$-6.56$                &-      &-      &LIME       &-          &-\\
                    
            AzV 96  &$21.7$                 &$2.66$                 &$5.41$                 &$-7.43$                &$0.1$  &$2.6$  &This study &UV+OPT     &C\\
                    &$21.3^{+1.6}_{-1.2}$    &$2.31^{+0.3}_{-0.2}$    &$5.23^{+0.1}_{-0.1}$    &$-6.81^{+0.2}_{-0.3}$   &$0.1$  &$1.0$  &JB+25      &OPT        &F\\
                    &-                      &-                      &-                      &$-6.99$                &-      &-      &LIME       &-          &-\\
            
            AzV 264 &$21.1$                 &$2.51$                 &$5.44$                 &$-7.32$                &$0.1$  &$3.0$  &This study &UV+OPT     &C\\
                    &$21.3^{+0.8}_{-1.2}$    &$2.31^{+0.1}_{-0.1}$    &$5.3^{+0.1}_{-0.1}$     &$-6.88^{+0.2}_{-0.3}$   &$0.1$  &$1.0$  &JB+25      &OPT        &F\\
                    &$21.5$                 &$2.5$                  &$5.43$                 &$-6.66$                &$0.2$  &$1.8$  &BP+24      &UV+OPT     &C\\
                    &$22.5^{+1.5}_{-1.5}$    &$2.55^{+0.2}_{-0.2}$    &$5.44$                 &$-6.54$                &$1$    &$2.5$  &T+05       &OPT        &F\\
                    &-                      &-                      &-                      &$-6.92$                &-      &-      &LIME       &-          &-\\
            
            AzV 175 &$20.0$                 &$2.80$                 &$4.97$                 &$-7.55$                &$0.2$  &$1.0$  &This study &UV+OPT     &C\\
                    &$19.0^{+1.6}_{-0.8}$    &$2.5^{+0.3}_{-0.1}$     &$4.84^{+0.1}_{-0.1}$    &$-8.52^{+0.5}_{-1.1}$   &$0.1$  &$1.0$  &JB+25      &OPT        &F\\
                    &-                      &-                      &-                      &$-7.19$                &-      &-      &LIME       &-          &-\\
                  \hline
\end{tabular}
\tablebib{JB+25: \citep{Bestenlehner2025}. FB+24: \citep{backs2024}, BP+24: \citet{bernini2024}, T+05: \citep{trundle2005}, T+04: \citep{trundle2004}}
\tablefoot{Obtained using F: \textsc{FASTWIND}, C: \textsc{CMFGEN}. The mass-loss rates are not corrected for clumping.}
\end{threeparttable}
\end{table*}

\begin{table*}[!h]
    \ContinuedFloat
  \centering   
\begin{threeparttable}[b]
    \caption{Continued.}              
        \def\arraystretch{1.2}                            
        \small                               
        \addtolength{\tabcolsep}{0.5em}
        \begin{tabular}{c c c c c c c c c c}          
            \hline
            Target  &$T_{\rm eff}$          &$\log{g}$              &$\log{L}$              &$\log{\dot{M}}$            &$f_{\rm vol,\infty}$  &$\beta$  &Ref.       &wavelength &Code \\[1.50 pt]
            
                    &kK                     &${\rm cm\,s^{-2}}$     &$L_{\odot}$            &$M_{\odot}\,{\rm yr}^{-1}$ &   &       &           &       &\\   
            \hline                        
            Sk 191  &$21.1$                 &$2.47$                 &$5.67$                 &$-6.81$                &$0.2$  &$2.5$  &This study &UV+OPT     &C\\
                    &$22.5^{+0.8}_{-1.2}$    &$2.5^{+0.1}_{-0.1}$     &$5.79^{+0.1}_{-0.1}$    &$-6.34^{+0.2}_{-0.2}$   &$0.1$  &$1.0$  &JB+25      &OPT        &F\\
                    &$19.0$                 &$2.25$                 &$5.67$                 &$-6.02$                &$0.8$  &$2.8$  &BP+24      &UV+OPT     &C\\
                    &$22.5^{1.5}_{-1.5}$    &$2.55^{0.2}_{-0.2}$    &$5.77$                 &$-6.17$                &$1$    &$2.0$  &T+04       &OPT        &F\\
                    &-                      &-                      &-                      &$-6.62$                &-      &-      &LIME       &-          &-\\
                    
            AzV 18    &$18.5$                 &$2.27$                 &$5.45$                 &$-7.19$                &$0.1$  &$3.5$  &This study &UV+OPT     &C\\
                    &$23.7^{+1.2}_{-2.0}$    &$2.69^{+0.1}_{-0.3}$    &$5.75^{+0.1}_{-0.1}$    &$-6.63^{+0.2}_{-0.2}$   &$0.1$  &$1.0$  &JB+25      &OPT        &F\\
                    &$19.0$                 &$2.35$                 &$5.45$                 &$-6.66$                &$1.5$  &$1.0$  &BP+24      &UV+OPT     &C\\
                    &$19.0^{+2.0}_{-2.0}$    &$2.3^{+0.2}_{-0.2}$     &$5.44$                 &$-6.64$                &$1$    &$3.0$  &T+04       &OPT        &F\\
                    &-                      &-                      &-                      &$-6.6$                 &-      &-      &LIME       &-          &-\\
            
    NGC330-ELS-04   &$17.1$                 &$2.36$                 &$4.93$                 &$-7.71$                &$0.2$  &$1.0$  &This study &UV+OPT     &C\\
                    &$17.8^{+2.0}_{-0.8}$    &$2.31^{+0.3}_{-0.2}$    &$4.96^{+0.2}_{-0.1}$    &$-7.77^{+0.3}_{-0.2}$   &$0.1$  &$1.0$  &JB+25      &OPT        &F\\
                    &-                      &-                      &-                      &$-7.31$                &-      &-      &LIME       &-          &-\\
            
            AzV 187   &$16.8$                 &$2.21$                 &$5.37$                 &$-7.77$                &$0.2$  &$1.0$  &This study &UV+OPT     &C\\
                    &$16.5$                 &$2.2$                  &$5.33$                 &$-7.22$                &$0.3$  &$2.2$  &BP+24      &UV+OPT     &C\\
                    &-                      &-                      &-                      &$-6.84$                &-      &-      &LIME       &-          &-\\
            
            AzV 22    &$14.2$                 &$1.81$                 &$5.22$                 &$-6.75$                &$0.1$  &$2.5$  &This study &UV+OPT     &C\\
                    &$15.9^{+0.8}_{-0.8}$    &$2.12^{+0.1}_{-0.6}$    &$5.28^{+0.1}_{-0.1}$    &$-7.38^{+0.3}_{-0.2}$   &$0.1$  &$1.0$  &JB+25      &OPT        &F\\
                    &$14.0$                 &$1.9$                  &$5.27$                 &$-7.15$                &$0.1$  &$3.7$  &BP+24      &UV+OPT     &C\\
                    &$14.5^{+1.5}_{-1.5}$    &$1.9^{+0.2}_{-0.2}$     &$5.04$                 &$-6.64$                &$1$    &$1.0$  &T+04       &OPT        &F\\
                    &-                      &-                      &-                      &$-7.6$                 &-      &-      &LIME       &-          &-\\
            
            AzV 393   &$13.3$                 &$1.61$                 &$5.53$                 &$-5.94$                &$0.1$  &$2.5$  &This study &UV+OPT     &C\\
                    &-                      &-                      &-                      &$-7.51$                &-      &-      &LIME       &-          &-\\
            
            AzV 343   &$12.0$                 &$2.00$                 &$4.65$                 &$-8.2$                 &$0.1$  &$1.0$  &This study &UV+OPT     &C\\
                    &$12.7^{+0.4}_{-0.4}$    &$1.88^{+0.2}_{-0.1}$    &$4.67^{+0.1}_{-0.1}$    &$-8.17^{+0.4}_{-0.9}$   &$0.1$  &$1$    &JB+25      &OPT        &F\\
                    &$12.5$                 &$2.05$                 &$4.72$                 &$-8.0$                 &$1.0$  &$2.0$  &BP+24      &UV+OPT     &C\\
                    &-                      &-                      &-                      &$-8.83$                &-      &-      &LIME       &-          &-\\
             
            AzV 324   &$12.5$                 &$2.21$                 &$4.70$                 &$-8.05$                &$0.1$  &$1.0$  &This Study &UV+OPT     &C\\ 
                    &$12.7^{+0.4}_{-0.4}$    &$1.88^{+0.2}_{-0.3}$    &$4.65^{+0.1}_{-0.1}$    &$-8.67^{+0.7}_{-0.6}$   &$0.1$  &$1$    &JB+25      &OPT        &F\\
                    &-                      &-                      &-                      &$-8.78$                &-      &-      &LIME       &-          &-\\
                  \hline
\end{tabular}
\tablebib{JB+25: \citep{Bestenlehner2025}. FB+24: \citep{backs2024}, BP+24: \citet{bernini2024}, T+05: \citep{trundle2005}, T+04: \citep{trundle2004}}
\tablefoot{Obtained using F: \textsc{FASTWIND}, C: \textsc{CMFGEN}. The mass-loss rates are not corrected for clumping.}
\end{threeparttable}
\end{table*}
\FloatBarrier
\clearpage

\end{appendix}
\end{document}